\documentclass[aps,prd,twocolumn,groupedaddress,amsmath,amssymb]{revtex4-2}

\usepackage{graphicx}
\usepackage{dcolumn}
\usepackage{bm}
\usepackage{rotating}
\usepackage{lineno}
\usepackage{subfigure}
\usepackage{lipsum}
\usepackage{epsfig}
\usepackage{overpic}
\usepackage{booktabs}
\usepackage{enumerate}
\usepackage{comment}
\usepackage[colorlinks,
            urlcolor=blue,
            linkcolor=blue,
            anchorcolor=blue,
            citecolor=blue]{hyperref}

\newcommand{\nJpsi}{1.0087} \newcommand{\nJpsiErr}{0.0044} \newcommand{\nJpsiErrR}{0.44} \newcommand{\nJpsiIncMC}{1.0011\times10^{10}}
\newcommand{\nBkgMuMu}{46.7\pm2.5} \newcommand{\nBkgEE}{1028.9\pm24.1}
\newcommand{\nObsMuMu}{695.0\pm26.4_{\rm stat.}} \newcommand{\nObsEE}{22803.0\pm151.0_{\rm stat.}}
\newcommand{\brMuMu}{2.97} \newcommand{\brMuMuErrSta}{0.11_{\rm stat.}} \newcommand{\brMuMuErrSys}{0.07_{\rm sys.}}
\newcommand{\ffMuMu}{1.645} \newcommand{\ffMuMuErrSta}{0.343_{\rm stat.}} \newcommand{\ffMuMuErrSys}{0.017_{\rm sys.}}
\newcommand{\brEE}{6.79} \newcommand{\brEEErrSta}{0.04_{\rm stat.}} \newcommand{\brEEErrSys}{0.36_{\rm sys.}}
\newcommand{\ffEE}{1.668} \newcommand{\ffEEErrSta}{0.093_{\rm stat.}} \newcommand{\ffEEErrSys}{0.024_{\rm sys.}}
\newcommand{\brEEComb}{6.93} \newcommand{\brEEErrTotComb}{0.28_{\rm tot.}}
\newcommand{\ffEEComb}{1.707} \newcommand{\ffEEErrStaComb}{0.076_{\rm stat.}} \newcommand{\ffEEErrSysComb}{0.029_{\rm sys.}}

\newcommand{\sysUFFGamMuMu}{0.3} \newcommand{\sysUFFGamEE}{0.1} \newcommand{\sysUFFTrkMuMu}{0.4} \newcommand{\sysUFFTrkEE}{1.2} \newcommand{\sysUFFPIDMuMu}{0.4} \newcommand{\sysUFFPIDEE}{0.3} \newcommand{\sysUFFKmfitMuMu}{0.5} \newcommand{\sysUFFKmfitEE}{0.1}   \newcommand{\sysUFFBkgMuMu}{0.6} \newcommand{\sysUFFBkgEE}{0.4} \newcommand{\sysUFFGamConvEE}{0.5}
\newcommand{\sysUBFGamMuMu}{0.7} \newcommand{\sysUBFGamEE}{0.7} \newcommand{\sysUBFTrkMuMu}{1.1} \newcommand{\sysUBFTrkEE}{4.7} \newcommand{\sysUBFPIDMuMu}{0.4} \newcommand{\sysUBFPIDEE}{1.1} \newcommand{\sysUBFKmfitMuMu}{0.6} \newcommand{\sysUBFKmfitEE}{0.1}  \newcommand{\sysUBFModelEE}{0.1} \newcommand{\sysUBFBkgMuMu}{0.3} \newcommand{\sysUBFBkgEE}{0.1} \newcommand{\sysUBFGamConvEE}{0.1}
\newcommand{\sysUBFTotalMuMu}{2.4} \newcommand{\sysUBFTotalEE}{5.2} \newcommand{\sysUFFTotalMuMu}{1.0} \newcommand{\sysUFFTotalEE}{1.5} \newcommand{\sysUBFTotalEEComb}{4.0} \newcommand{\sysUFFTotalEEComb}{1.8}

\let\oldequation\equation
\let\oldendequation\endequation
\renewenvironment{equation}{\linenomathNonumbers\oldequation}{\oldendequation\endlinenomath}

\begin{document}

\title{Measurement of the $\eta$ transition form factor through $\eta' \rightarrow \pi^+\pi^-\eta$ decay}

\author{
\begin{small}
  \begin{center}
M.~Ablikim$^{1}$, M.~N.~Achasov$^{4,c}$, P.~Adlarson$^{76}$, X.~C.~Ai$^{81}$, R.~Aliberti$^{35}$, A.~Amoroso$^{75A,75C}$, Q.~An$^{72,58,a}$, Y.~Bai$^{57}$, O.~Bakina$^{36}$, Y.~Ban$^{46,h}$
% ...
\\
\vspace{0.2cm}
(BESIII Collaboration)\\
\vspace{0.2cm} {\it
$^{1}$ Institute of High Energy Physics, Beijing 100049, People's Republic of China\\
% ...
$^{4}$ Budker Institute of Nuclear Physics SB RAS (BINP), Novosibirsk 630090, Russia\\
% ...
$^{35}$ Johannes Gutenberg University of Mainz, Johann-Joachim-Becher-Weg 45, D-55099 Mainz, Germany\\
$^{36}$ Joint Institute for Nuclear Research, 141980 Dubna, Moscow region, Russia\\
% ...
$^{46}$ Peking University, Beijing 100871, People's Republic of China\\
% ...
$^{57}$ Southeast University, Nanjing 211100, People's Republic of China\\
$^{58}$ State Key Laboratory of Particle Detection and Electronics, Beijing 100049, Hefei 230026, People's Republic of China\\
% ...
$^{72}$ University of Science and Technology of China, Hefei 230026, People's Republic of China\\
% ...
$^{75}$ University of Turin and INFN, (A)University of Turin, I-10125, Turin, Italy; (B)University of Eastern Piedmont, I-15121, Alessandria, Italy; (C)INFN, I-10125, Turin, Italy\\
$^{76}$ Uppsala University, Box 516, SE-75120 Uppsala, Sweden\\
% ...
$^{81}$ Zhengzhou University, Zhengzhou 450001, People's Republic of China\\

\vspace{0.2cm}
$^{a}$ Deceased\\
% ...
$^{c}$ Also at the Novosibirsk State University, Novosibirsk, 630090, Russia\\
% ...
$^{h}$ Also at State Key Laboratory of Nuclear Physics and Technology, Peking University, Beijing 100871, People's Republic of China\\
% ...

}

\end{center}
\end{small}
}

\begin{abstract}
Based on a sample of $(\nJpsi\pm\nJpsiErr)\times10^{10}$ $J/\psi$ events collected at BESIII, the transition form factor of the $\eta$ meson is extracted by analyzing $J/\psi\to\gamma\eta',~\eta'\to\pi^+\pi^-\eta,~\eta\to\gamma l^+l^-$ ($l$=$e$, $\mu$) events. The measured slope of the transition form factor is $\Lambda^{-2}=\ffEE\pm\ffEEErrSta\pm\ffEEErrSys$ (GeV/$c^2$)$^{-2}$ for the di-electron channel and $\Lambda^{-2}=\ffMuMu\pm\ffMuMuErrSta\pm\ffMuMuErrSys$ (GeV/$c^2$)$^{-2}$ for the di-muon channel. 
The branching fractions for $\eta\rightarrow\gamma e^+e^-$ and $\eta\rightarrow\gamma\mu^+\mu^-$ are measured to be $\mathcal{B}(\eta\to\gamma e^+e^-)=(\brEE\pm\brEEErrSta\pm\brEEErrSys)\times 10^{-3}$ and $\mathcal{B}(\eta\to\gamma\mu^+\mu^-)=(\brMuMu\pm\brMuMuErrSta\pm\brMuMuErrSys)\times 10^{-4}$. 
By combining with the results based on the $J/\psi\to\gamma\eta,~\eta\to\gamma e^+e^-$ events from the previous BESIII measurement, we determine $\Lambda^{-2}=\ffEEComb\pm\ffEEErrStaComb\pm\ffEEErrSysComb$ (GeV/$c^2$)$^{-2}$ and $\mathcal{B}(\eta\to\gamma e^+e^-)=(\brEEComb\pm\brEEErrTotComb)\times 10^{-3}$. In addition, we search for the dark photon ($A'$) using the combined events. No significant signal is observed, and the upper limits on $\mathcal{B}(\eta\to\gamma A',~A'\to e^+e^-)$ are set at 90\% confidence level for different $A'$ mass hypotheses.
\end{abstract}

\maketitle

\section{Introduction}
Transition form factors (TFF) provide insight into the internal structure of hadrons, including how charge and magnetization are distributed among their constituents. Understanding TFF is crucial for comprehending the binding and confinement of quarks and gluons in hadrons, which is a fundamental concept in Quantum Chromodynamics. Additionally, the TFF of light mesons has garnered recent attention due to their contribution in calculating the anomalous magnetic moment of the muon ($a_{\mu}$)~\cite{Aoyama:2020ynm}.

In this study, the TFF of the $\eta$ meson is determined through the $\eta\to\gamma l^+l^-$ ($l=e$, $\mu$) decays, where the lepton pair is formed by internal conversion of an intermediate virtual photon. In the vector meson dominance (VMD) model, the interactions between a virtual photon and hadrons are assumed to be dominated by a superposition of neutral vector meson states. The TFF can be parameterized as~\cite{Landsberg:1985gaz}
\begin{equation}
	F(q^2)=N\sum_V\frac{g_{\eta\gamma V}}{2g_{V\gamma}}\frac{m^2_V}{m^2_V-q^2-i\Gamma_Vm_V},
\end{equation}
where $q^2$ represents the squared invariant mass of the lepton pair; $N$ is a normalization constant ensuring $F(0)=1$; $V$ stands for vector mesons such as $\rho$, $\omega$, and $\phi$; $m_V$ and $\Gamma_V$ are the mass and width of $V$, and $g_{\eta\gamma V}$ and $g_{V\gamma}$ correspond to the respective coupling constants. When there is a single dominant vector meson, the single-pole approximation is often used:
\begin{equation}
	F(q^2)=\frac{1}{1-q^2/\Lambda^2}.\label{eq_TFF}
\end{equation}

Here, the single parameter $\Lambda$ can be experimentally determined as the slope of the TFF, defined as:
\begin{equation}
	{\rm slope}\equiv\frac{{\rm d}F}{{\rm d}q^2}\bigg|_{q^2=0}=\Lambda^{-2}.\label{eq_slope}
\end{equation}

Before this work, three measurements of the TFF of the $\eta$ meson have been performed by the A2 collaboration~\cite{Adlarson:2016hpp} through the $\eta\to\gamma e^+e^-$ channel, the NA60 collaboration~\cite{NA60:2016nad} through the $\eta\to\gamma\mu^+\mu^-$ channel, and the {\sc BESIII} collaboration~\cite{BESIII:2024pxo} through the $\eta\to\gamma e^+e^-$ channel. 

In contrast with the previous BESIII work~\cite{BESIII:2024pxo}, where $\eta$ candidates have been obtained through the $J/\psi\rightarrow\gamma\eta$ decay (Sample II), we use $\eta$ mesons produced by the decay $\eta'\rightarrow\pi^+\pi^-\eta$, where the $\eta'$ comes from the radiative decay of the $J/\psi$, namely $J/\psi\rightarrow\gamma\eta'$ (Sample I). This analysis is based on a data sample of $(\nJpsi\pm\nJpsiErr)\times10^{10}$ $J/\psi$ events collected with the BESIII detector during 2009-2019~\cite{BESIII:2021cxx}.
Due to the better mass resolution of the $\eta'$, the new approach benefits from reduced backgrounds from the $J/\psi$ decays. Besides, Sample I has access to more signal events due to the higher branching fraction (BF), and its reconstruction efficiency is about two times larger than that of Sample II after taking into account the tracking efficiency of the two charged pions. In this work, the TFF of $\eta$ is measured with Sample I using a new approach at first, then we re-measure it using the combined Sample I and II. The BFs of $\eta\to\gamma l^+l^-$ are also measured.

Additionally, new light hidden particles, such as axion-like particles and the dark photons, which may couple to light quarks or gluons, could be produced in the Dalitz decay $\eta\rightarrow\gamma e^+e^-$~\cite{Gan:2020aco}. 
We have also searched for dark photons by using the samples mentioned above. 

\section{Detector and data samples}
The BESIII detector~\cite{BESIII:2009fln} records symmetric $e^+e^-$ collisions provided by the BEPCII storage ring~\cite{Yu:2016cof} in the center-of-mass energy range from 1.84 to 4.95~GeV, with a peak luminosity of $1.2 \times 10^{33}\;\text{cm}^{-2}\text{s}^{-1}$  achieved at $\sqrt{s} = 3.773\;\text{GeV}$. BESIII has collected large data samples in this energy region~\cite{BESIII:2020nme, Jiao:2020dqs, Zhang:2022bdc}. The cylindrical core of the BESIII detector covers 93\% of the full solid angle and consists of a helium-based multilayer drift chamber~(MDC), a plastic scintillator time-of-flight system~(TOF), and a CsI(Tl) electromagnetic calorimeter~(EMC), which are all enclosed in a superconducting solenoidal magnet providing a 1.0~T magnetic field. The magnetic field was 0.9~T in 2012. The solenoid is supported by an octagonal flux-return yoke with resistive plate counter muon identification modules interleaved with steel.  
The charged-particle momentum resolution at $1~{\rm GeV}/c$ is $0.5\%$, and the ${\rm d}E/{\rm d}x$ resolution is $6\%$ for electrons from Bhabha scattering. The EMC measures photon energies with a resolution of $2.5\%$ ($5\%$) at $1$~GeV in the barrel (end cap) region. The time resolution in the TOF barrel region is 68~ps, while that in the end cap region was 110~ps. The end cap TOF system was upgraded in 2015 using multigap resistive plate chamber technology, providing a time resolution of 60~ps~\cite{Cao:2020ibk}, which benefits 87\% of the data used in this analysis.

Simulated data samples produced with the {\sc geant4}-based~\cite{GEANT4:2002zbu} Monte Carlo (MC) package~\cite{MCPackage}, which includes the geometric description of the BESIII detector and the detector response~\cite{You_2008, Liang:2009zzb}, are used to determine the detection efficiency and estimate the backgrounds. The simulation includes the beam energy spread and initial state radiation in the $e^+e^-$ annihilations modeled with the generator {\sc kkmc}~\cite{Jadach:2000ir}. A sample of $\nJpsiIncMC$ simulated inclusive $J/\psi$ events is used to estimate the background events. The inclusive MC sample includes both the production of the $J/\psi$ resonance and the continuum processes incorporated in {\sc kkmc}. The known decay modes are modeled with {\sc evtgen}~\cite{Ping:2008zz} using BFs taken from the Particle Data Group (PDG)~\cite{ParticleDataGroup:2024cfk}, and the remaining unknown charmonium decays are modeled with {\sc lundcharm}~\cite{Chen:2000tv, Yang:2014vra}. Final state radiation from charged final state particles is incorporated using {\sc photos}~\cite{Richter-Was:1992hxq}.

In addition, exclusive MC samples are generated to determine the detection efficiency and study the background distributions. The simulated processes and the corresponding generator models are listed in Table~\ref{table_generators}.
\begin{table}[hb]
	\centering
	\caption{Generator models used for MC simulations.}\label{table_generators}
	\begin{tabular}{lc}\hline\hline
		Decay mode & Generator model\\\hline
		$J/\psi\to\gamma\eta'$ & Helicity amplitude~\cite{Morisita:1990cg}\\
		$\eta'\to\pi^+\pi^-\eta$ & Dalitz plot analyses~\cite{BESIII:2017djm}\\
		$\eta\to\gamma\mu^+\mu^-$ & Transition form factor~\cite{Dzhelyadin:1980kh}\\
		$\eta\to\gamma e^+e^-$ & Transition form factor~\cite{Berghauser:2011zz}\\
		$\eta\to\gamma\pi^+\pi^-$ & Partial wave analysis~\cite{BESIII:2017kyd}\\
		$\eta'\to\pi^+\pi^-e^+e^-$ & VMD model~\cite{Zhang:2012gt, BESIII:2024awu}\\
		$\eta\to\gamma\gamma$ & Phase space\\
        $\eta\to\gamma A'$ & Phase space \\
        $A'\to e^+e^-$ & Phase space \\
		\hline\hline
	\end{tabular}
\end{table}

\section{Study of Sample I}
In this section, the Sample I is selected, and the TFF of the $\eta$ meson and the BFs of the $\eta\to\gamma l^+l^-$ ($l$=$e$, $\mu$) decays are measured using this sample.

\subsection{Event Selection and Background Analysis}\label{sec_evtSelAndBkgAna}
Candidate events for $J/\psi\rightarrow\gamma \eta',~\eta'\rightarrow\pi^+\pi^-\eta,~\eta\rightarrow\gamma l^+ l^-(l=e,~\mu)$ are subjected to several selection criteria. Firstly, at least two photons must be reconstructed using information from the EMC. To suppress fake photon candidates, the deposited energy of each EMC shower must be more than 25~MeV in the barrel region ($|\cos\theta|<$ 0.8) and more than 50~MeV in the end cap region (0.86 $<|\cos\theta|<$ 0.92), where $\theta$ is the polar angle defined with respect to the $z$-axis, which is the symmetry axis of the MDC. The opening angle between the detected position of the photon candidate and the closest extrapolated charged track must be larger than 10 degrees to exclude showers originating from charged tracks. Additionally, the difference between the EMC time of the photon candidate and the event start time must be within $[0, 700]~{\rm ns}$ to suppress electronic noise and unrelated photons. Furthermore, candidate events must have two positively and two negatively charged tracks reconstructed using information from the MDC. These tracks are required to be within a polar angle range of $|\rm{cos\theta}|<0.93$, and to pass within $10$ cm of the interaction point (IP) along the $z$-axis and within 1 cm in the transverse plane.

By utilizing information from the MDC ($dE/dx$), TOF, and EMC detectors, particle identification (PID) is applied to charged tracks, and we acquire the combined probabilities ($\rm{Prob}$) under the hypotheses of the track being an electron (positron), muon, or pion. 
For the $\eta\rightarrow\gamma \mu^+\mu^-$ channel, candidates are required to satisfy $\rm{Prob}(\pi)>\rm{Prob}(\mu)$ and $\rm{Prob}(\pi)>\rm{Prob}(e)$ for pions, and $\rm{Prob}(\mu)>\rm{Prob}(\pi)$ and $\rm{Prob}(\mu)>\rm{Prob}(e)$ for muons. For the $\eta\rightarrow\gamma e^+e^-$ channel, candidates are required to satisfy $\rm{Prob}(\pi)>\rm{Prob}(e)$ for pions, and $\rm{Prob}(e)>\rm{Prob}(\pi)$ for electrons (positrons). 
Events are retained only if they contain two oppositely charged pions and two oppositely charged leptons of the same flavor.

A kinematic fit to the final state particle candidates is performed, to adjust the particle momenta or energies within the measured uncertainties to satisfy the kinematic constraints. 
A six-constraint (6C) kinematic fit imposing energy-momentum conservation and constraints on the masses of the $\eta'$ and $\eta$ particles, taken from the PDG~\cite{ParticleDataGroup:2024cfk}, is performed. In the fit, the most energetic photon is considered as the radiative photon of the $J/\psi$ decay, denoted as $\gamma_r$. Another photon is supposed as the one from the $\eta$ decay, denoted as $\gamma_{\eta}$. If there are multiple $\gamma_{\eta}$ candidates, the one which minimizes the goodness of the fit $\chi^2_{\rm 6C}$  is retained for further analyses. 
Additionally, for the $\eta\to\gamma\mu^+\mu^-$ events we require $\chi^2_{\rm 6C}<$40, and for the $\eta\to\gamma e^+e^-$ events we require $\chi^2_{\rm 6C}<200$, by optimizing the Figure-of-Merit defined as $\mathit{S} / \sqrt{\mathit{S}+\mathit{B}}$, where $S$ is the number of events from the signal MC sample and $B$ is the number of background events from the inclusive MC sample.

\begin{figure}[htbp]
	\subfigure[]{\includegraphics[width=.8\columnwidth]{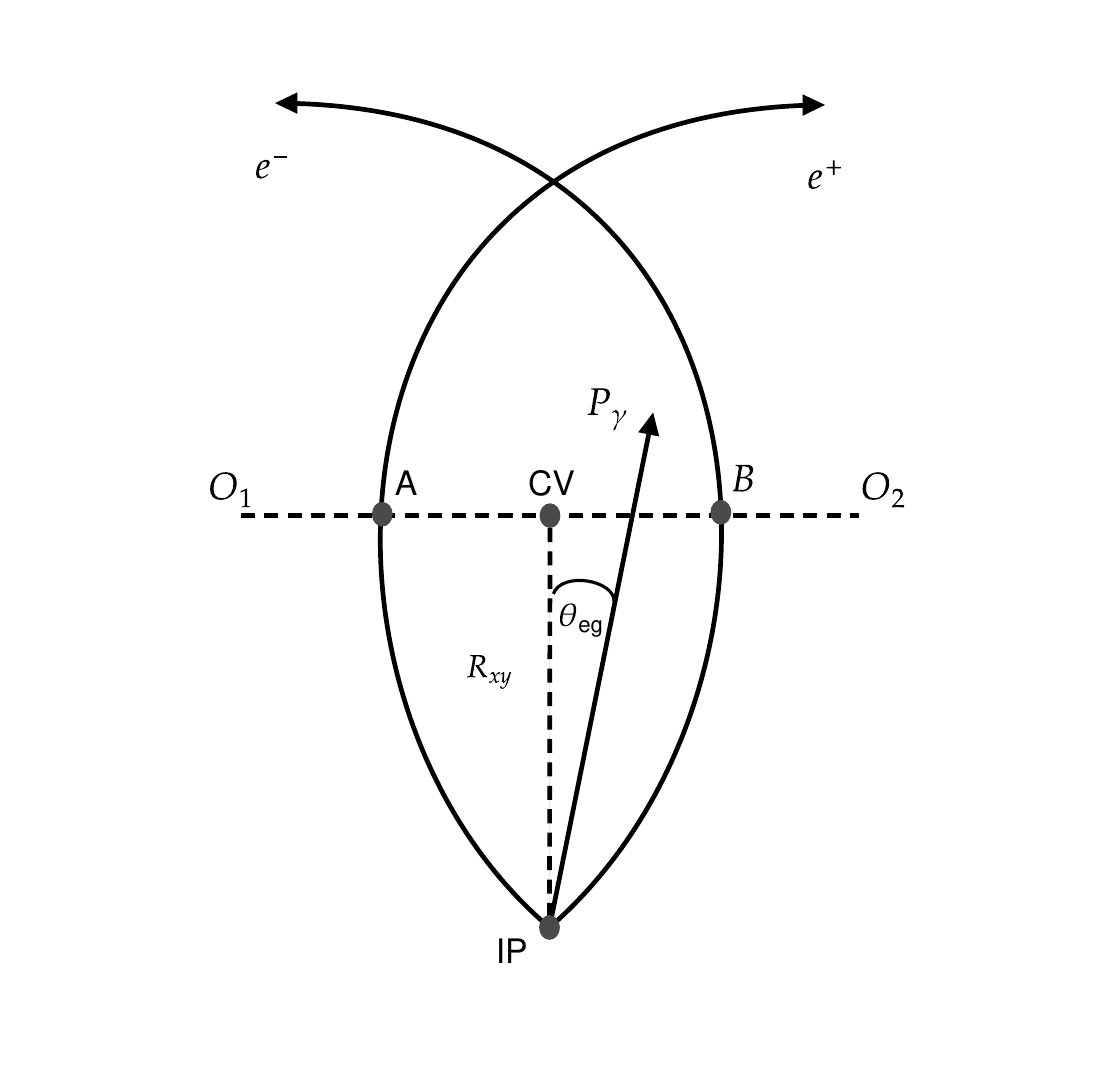}\label{fig_cutsGamConv-CP}}\\
	\subfigure[]{\includegraphics[width=.8\columnwidth]{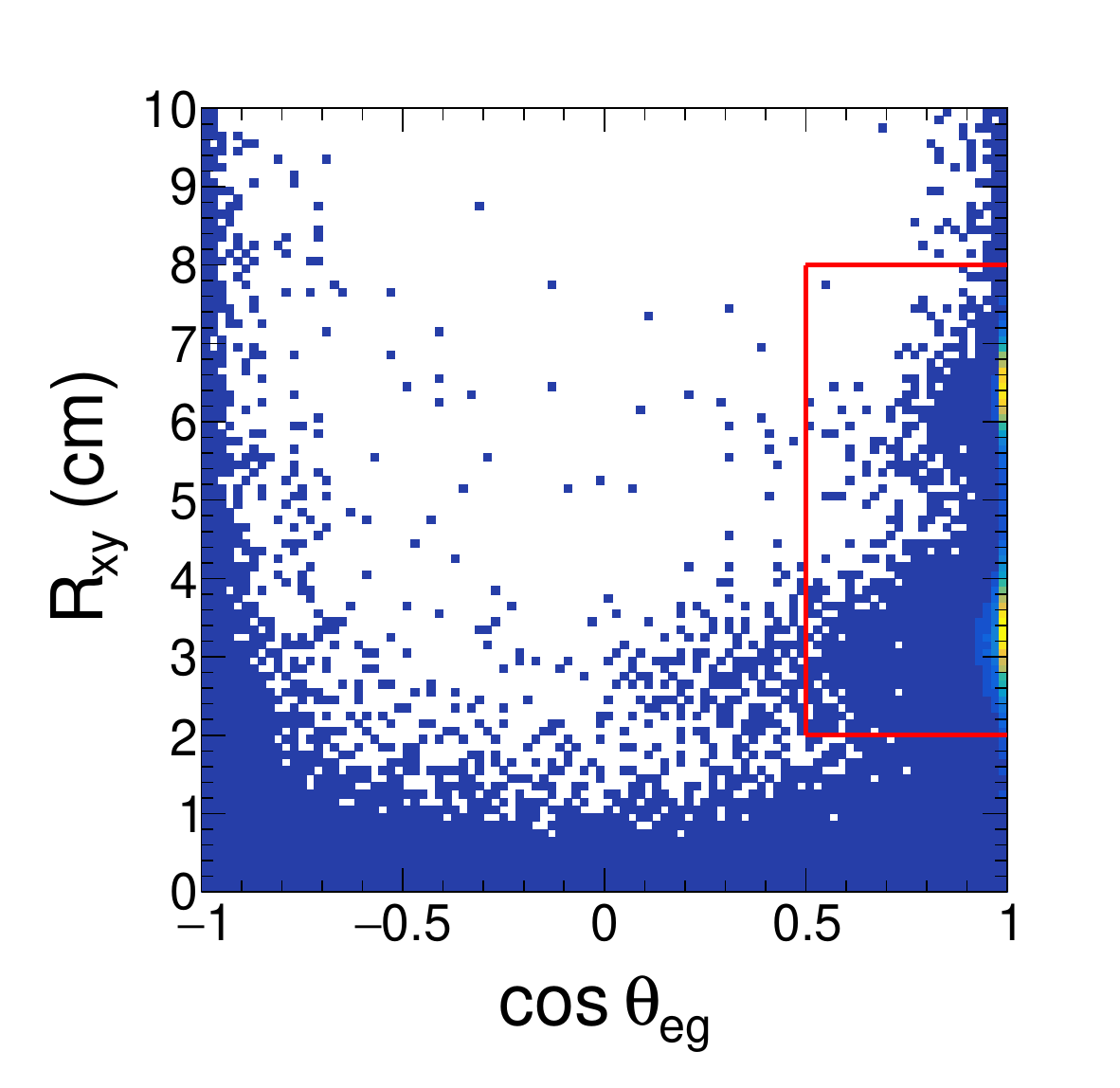}\label{fig_cutsGamConv-RxyCosEGData}}
	\caption{(a) Illustration of the CV reconstruction. The points $O_1$ and $O_2$ are the centers of the projections of the $e^+$ and $e^-$ tracks in the $x$-$y$ plane, respectively. The IP is the reconstructed vertex of the tracks. The CV obtained with the PCF is supposed to be the true vertex of the tracks. The distance from the IP to the CV is $R_{xy}$. The arrow \bm{$P$}$_{\gamma}$ represents the momentum of the converted photon, and the angle between the arrow and the IP-CV is $\theta_{eg}$. (b) Distribution of $R_{xy}$ versus cos$\theta_{eg}$ of the $\eta\to\gamma e^+e^-$ sample. Events in the red box are considered as the removed $\gamma$ conversion events.}
\end{figure}

After all the above criteria have been applied to select the $\eta\rightarrow\gamma l^+l^-$ ($l=e, \mu$) channel, the main source of background comes
from $\eta\to\gamma\gamma$ events, with one $\gamma$ converting in matter to an $e^+e^-$ pair. 
Converted photons can be reconstructed from the $e^+e^-$ pairs using the Photon Conversion Finder (PCF) package~\cite{Xu:2012xq}. At BESIII, the helix parameters of charged tracks are determined assuming that the IP is the origin, which is not true in $\gamma$ conversion case since the conversion vertex (CV) is generally displaced from the IP. In the PCF, the CV of the photon is estimated using the $e^+e^-$ track projections in the $x$-$y$ plane, perpendicular to the beam direction. The midpoint of the centers of the two track projections is taken as the CV, as shown in Fig.~\ref{fig_cutsGamConv-CP}. As most photon conversions occur at the beam pipe and at the inner wall of the MDC, which have higher material budget, the distances from the CV to the IP in the $x$-$y$ plane, denoted by $R_{xy}$, are usually greater than 2~cm. Moreover, the angle between the momentum vector of the converted photon and the direction from the IP to the CV, denoted by $\theta_{eg}$, is usually close to zero. Hence, events with 2 cm $< R_{xy} <$ 8 cm and cos$(\theta_{eg}) > $ 0.5 are rejected to suppress $\gamma$ conversion related background. The two-dimensional~(2D) distribution of $R_{xy}$ versus cos$\theta_{eg}$ of the $\eta\to\gamma e^+e^-$ sample is shown in Fig.~\ref{fig_cutsGamConv-RxyCosEGData}.

\begin{figure}[htbp]
    \centering
    \subfigure[]{\includegraphics[width=.9\columnwidth]{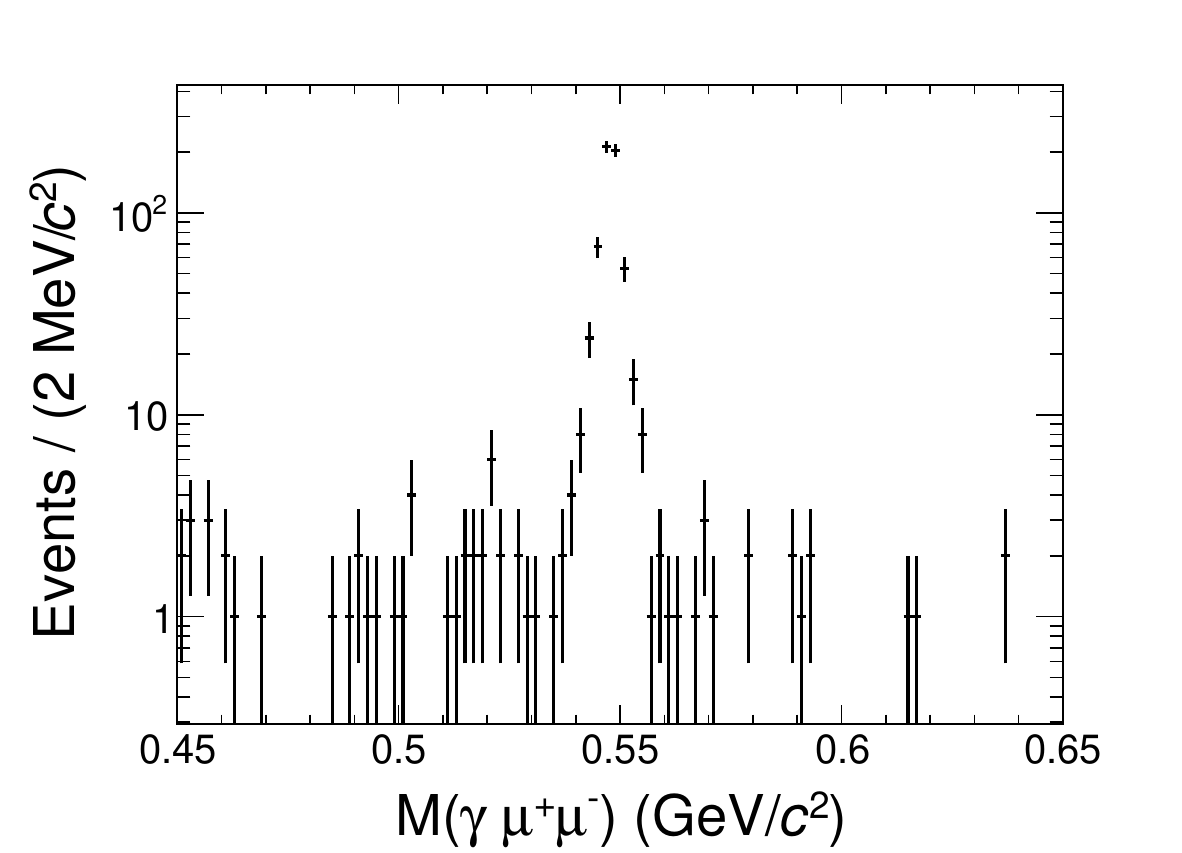}}\\
    \subfigure[]{\includegraphics[width=.9\columnwidth]{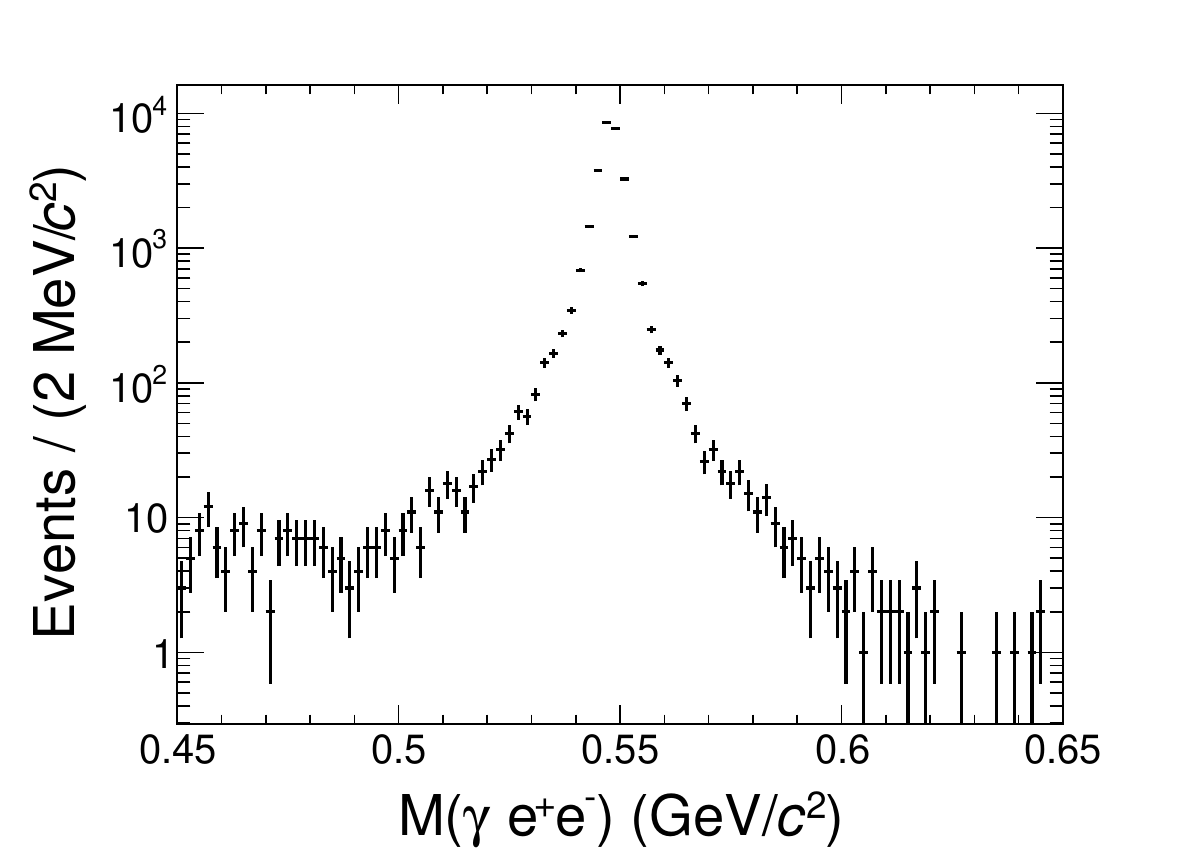}}
    \caption{Distributions of (a) $M(\gamma \mu^+\mu^-)$ and (b) $M(\gamma e^+e^-)$.}
    \label{fig:eta}
\end{figure}

After all event selections, the distributions of the $\gamma l^+l^-$ invariant masses  ($M(\gamma l^+l^-)$) of the accepted candidate events in data are shown in Fig.~\ref{fig:eta}.

Potential backgrounds from non-$J/\psi$ decay processes are estimated using the 169~pb$^{-1}$ of continuum data taken at $\sqrt{s}$=3.08 GeV. As no event passes all the selections, the non-$J/\psi$ decay backgrounds are ignored. The backgrounds from $J/\psi$ decays are studied with the inclusive MC sample. The three main backgrounds are studied using exclusively simulated MC samples;  we estimate the background yields using the known BFs, and the obtained results are shown in Tables~\ref{tab_bkgmm} and~\ref{tab_bkgee}.

\begin{table}[h]\centering
\caption{The estimated numbers of background events ($N_{\rm bkg}$) for different background sources for $\eta\to\gamma\mu^+\mu^-$.}\label{tab_bkgmm}
\begin{tabular}{c c}\hline\hline
	Background process  & $N_{\rm bkg}$\\\hline
	$J/\psi\to\gamma\eta',\eta'\to\pi^+\pi^-\eta,\eta\to\gamma\pi^+\pi^-$& 43.7$\pm2.2$\\
	$J/\psi\to\pi^+\pi^-\pi^+\pi^-\pi^0,\pi^0\to\gamma\gamma$ & 3.0$\pm0.3$\\
	\hline
	Total & $\nBkgMuMu$\\
	\hline\hline
\end{tabular}
\end{table}

\begin{table}[h]\centering
\caption{The estimated numbers of background events ($N_{\rm bkg}$) for different background sources for $\eta\to\gamma e^+e^-$.}\label{tab_bkgee}
\begin{tabular}{c c}\hline\hline
    Background process  & $N_{\rm bkg}$\\\hline
    $J/\psi\to\gamma\eta',\eta'\to\pi^+\pi^-\eta,\eta\to\gamma\gamma$&993.6$\pm15.7$\\
	$J/\psi\to\gamma\eta',\eta'\to\pi^+\pi^-e^+e^-$&0.3$\pm0.2$\\
	$J/\psi\to\gamma\pi^+\pi^-\eta,\eta\to\gamma e^+e^-$&25.0$\pm5.0$\\
	Other&10.0$\pm3.2$\\
	\hline
	Total&$\nBkgEE$\\
	\hline\hline
\end{tabular}
\end{table}

\subsection{Measurement of Form Factor}\label{sec_ff}
To extract the TFF of the $\eta$ meson, unbinned maximum likelihood fits are performed to the selected $\eta\to\gamma l^+l^-$ samples, with likelihood 
\begin{equation}
	\mathcal{L}=\prod^N_{i=1}{\cal P}(\xi_i,~\Lambda).
\end{equation}

Here, $N$ is the number of observed events, $\xi_i$ stands for the four-momenta of the final particles in the $i$-th event, $\mathcal{P}(\xi_i,~\Lambda)$ is the probability to observe the $i$-th event supposing $\Lambda$, calculated as:
\begin{equation}
    \mathcal{P}(\xi_{i},~\Lambda)=\frac{|\mathcal{A}(\xi_{i},~\Lambda)|^{2} \epsilon(\xi_{i})}{\int {\rm d}\xi |\mathcal{A}(\xi,~\Lambda)|^{2}\epsilon(\xi)},
\end{equation}
where $\epsilon(\xi_i)$ is the reconstruction efficiency of event $\xi_i$, $|\mathcal{A}(\xi_i,~\Lambda)|^2$ is the squared amplitude of the $\eta\to\gamma l^+l^-$ decay~\cite{Petri:2010ea}:
\begin{equation}
	|\mathcal{A}(\xi_i,~\Lambda)|^2=e^2\frac{(m_{\eta}^2-q^2)^2}{2q^2}(2-\beta^2{\rm sin}^2\theta)\mathcal{M}^{2}_{\eta}F^{2}(q^2),
\end{equation}
where $e$ is the electron charge constant, $m_{\eta}$ is the nominal mass of $\eta$, $q^2$ is $M^2(l^+l^-)$, $\beta=\sqrt{1-4m^2_{l^{\pm}}/q^2}$, $\theta$ is the angle between the $\gamma$ and the $l^{\pm}$ in the rest frame of $l^+l^-$, ${\cal M}_\eta$ is the pseudoscalar mesons mixing parameter~\cite{Petri:2010ea}, $F(q^2)$ is the form factor as defined in Eq.~(\ref{eq_TFF}).

The free parameters are estimated by \mbox{MINUIT}~\cite{James:1975dr}. The fit minimizes the negative log-likelihood value, calculated as:
\begin{equation}
\begin{split}
    -\ln{\mathcal{L}}&=\omega'[-\ln\mathcal{L}_{\rm data}-(-\ln\mathcal{L}_{\rm bkg})].
\end{split}
\end{equation}

The $\mathcal{L}_{\rm bkg}$ are estimated using the exclusive MC samples listed in Tables~\ref{tab_bkgmm} and~\ref{tab_bkgee}, and the numbers of background events are fixed.
To obtain an unbiased uncertainty estimation, the normalization factor derived from Ref.~\cite{Langenbruch:2019nwe} is defined as
\begin{equation}
    \omega^{\prime}=\frac{ N_{\rm data} - \sum_{j} N_{\rm bkg}^{j}\omega_{j} }{ N_{\rm data}+\sum_j N_{\rm bkg}^{j}\omega_{j}^{2} },
\end{equation}
where $N_{\rm data}$ and $N_{\rm bkg}^j$ are the numbers of events in data and in the $j$-th background process, and $\omega_{j}$ is the weight factor of the $j$-th background component. 

\begin{figure}[htbp]
\begin{center}
	\subfigure[]{\includegraphics[width=.9\columnwidth]{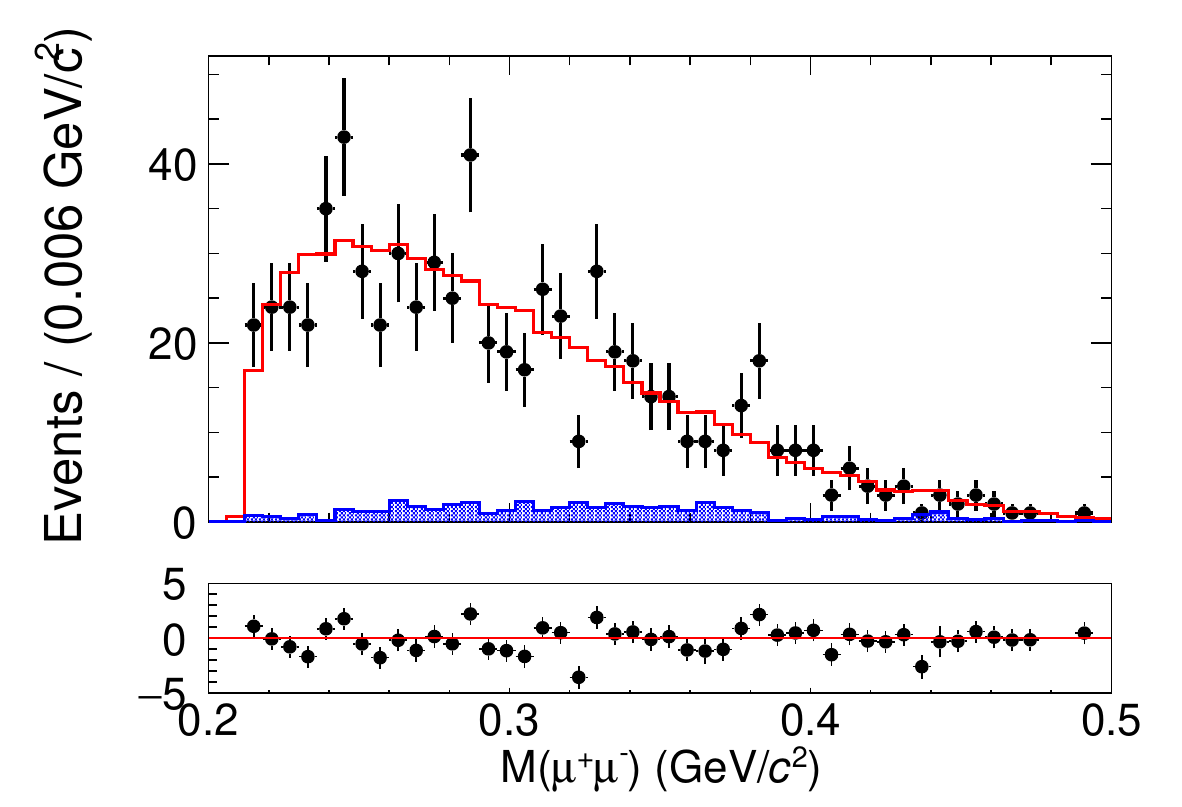}\label{fig_fitMuMu}}\\
	\subfigure[]{\includegraphics[width=.9\columnwidth]{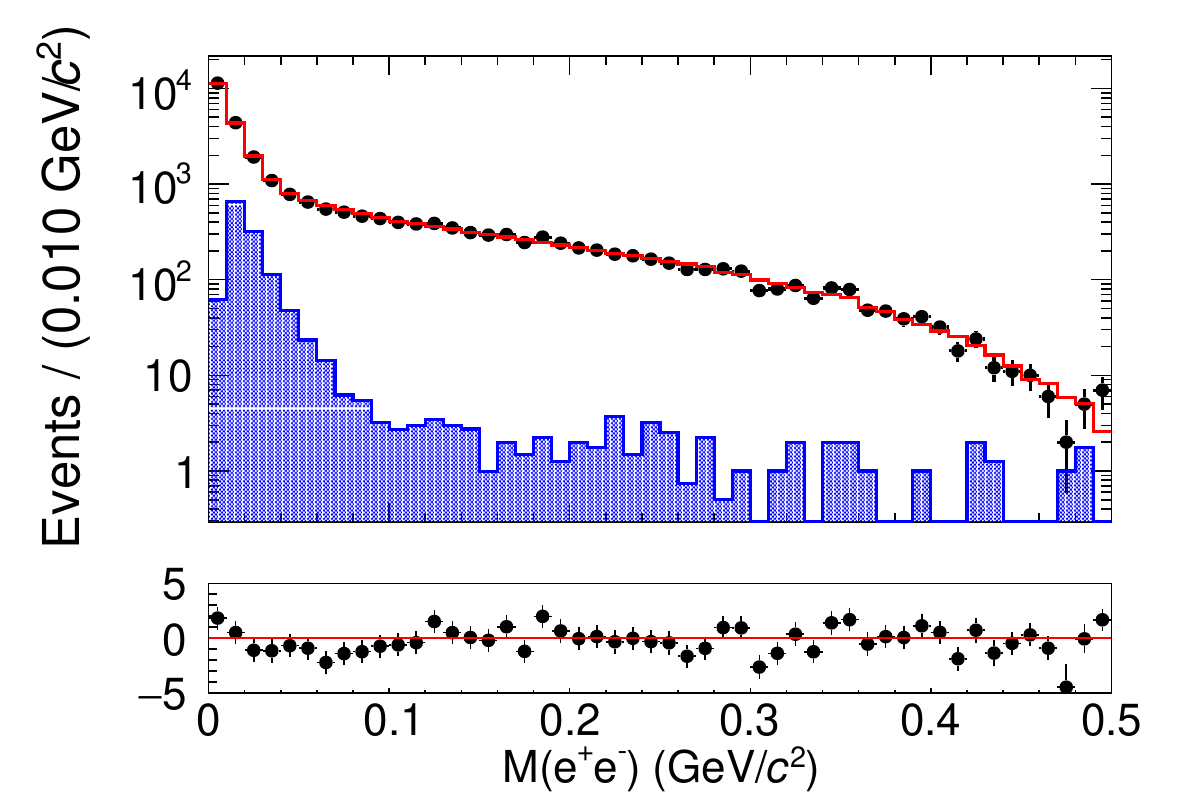}\label{fig_fitEE}}
	\caption{Invariant mass distributions of (a) $\mu^+\mu^-$ and (b) $e^+e^-$ pairs. The black dots with error bars are data, the red solid lines are the total fit results, including both the signal and the background, and the blue histograms are the background.}\label{fig_ff}
\end{center}
\end{figure}

The fits yield $\Lambda^{-2}(\eta\to\gamma e^+e^-)=\ffEE\pm\ffEEErrSta$ (GeV/$c^2$)$^{-2}$ and $\Lambda^{-2}(\eta\to\gamma\mu^+\mu^-)=\ffMuMu\pm\ffMuMuErrSta$ (GeV/$c^2$)$^{-2}$. The comparisons of the fit results and the distributions of data are shown in Fig.~\ref{fig_ff}.

\begin{figure*}[htbp]
\begin{center}
	\subfigure[]{\includegraphics[width=.9\columnwidth]{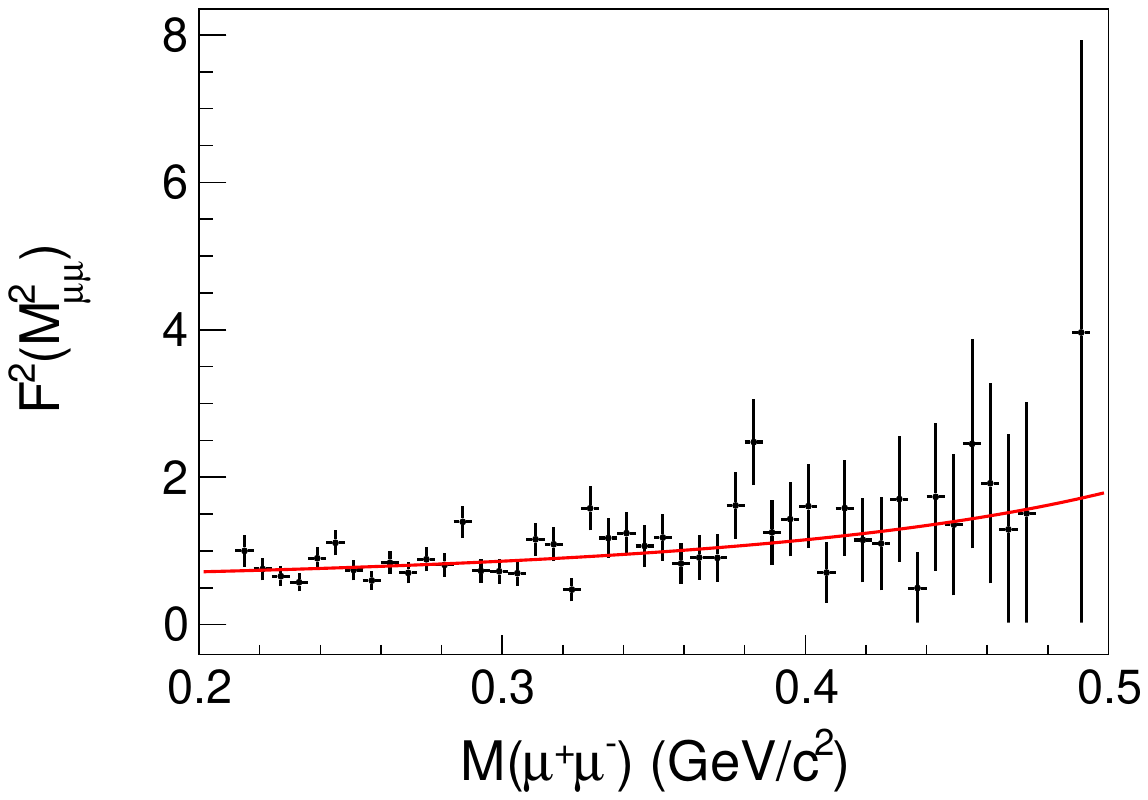}}
	\subfigure[]{\includegraphics[width=.9\columnwidth]{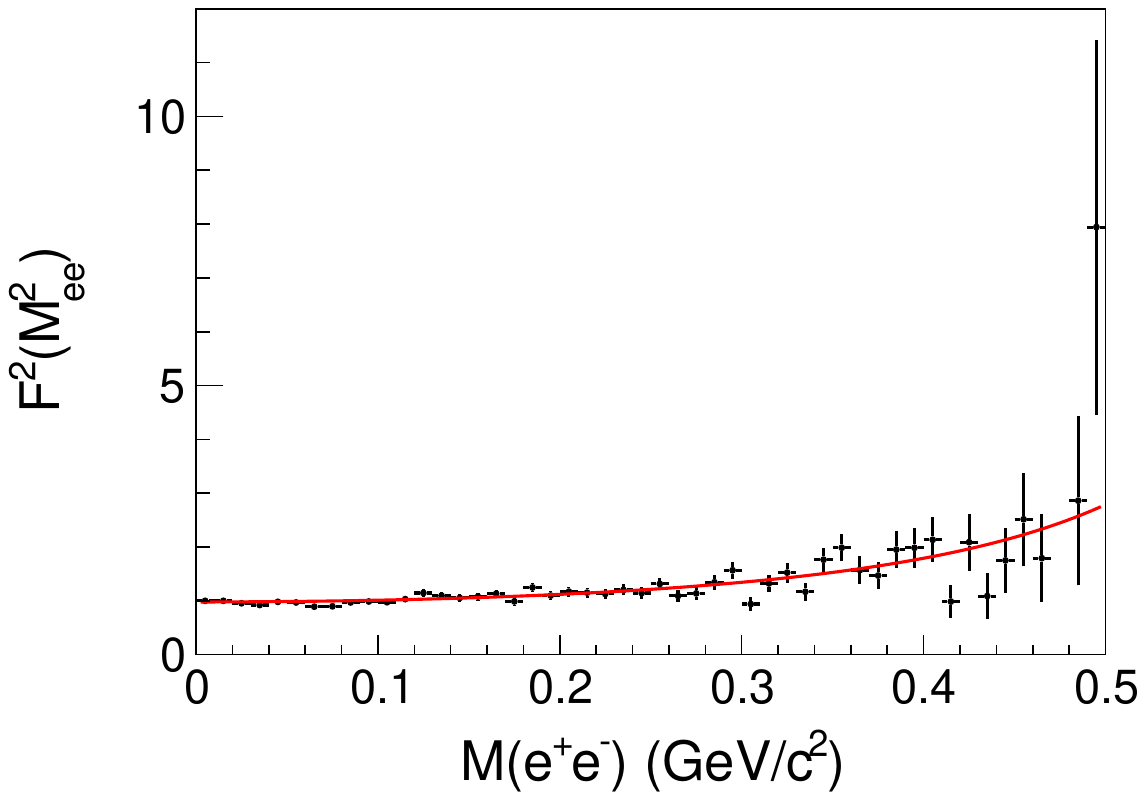}}
	\caption{The distributions of $F^2(q^2)$ over (a) $M(\mu^+\mu^-)$ and (b) $M(e^+e^-)$. The black dots with error bars are the ratios of the background-subtracted and efficiency-corrected data to the efficiency-corrected MC simulation using $F^2(q^2)\equiv 1$. The red lines are the $F^2(q^2)$ functions using the measured $\Lambda$ values.}\label{fig_ffdistribution}
\end{center}
\end{figure*}

To extract the TFF of $\eta$, we divide the background-subtracted and efficiency-corrected data to efficiency-corrected MC with $F^2(q^2)\equiv 1$, as shown in Fig.~\ref{fig_ffdistribution}.

\subsection{Measurements of Branching Fractions}
The BFs of the $\eta\to\gamma l^+l^-$ decays are calculated as:
\begin{equation}\label{eq_br}
    \mathcal{B}(\eta\to\gamma l^+l^-)=\frac{N_{\rm data}-N_{\rm bkg}}{N_{J/\psi}\cdot\epsilon\cdot\mathcal{B}(J/\psi\to\gamma\eta',\eta'\to\pi^+\pi^-\eta)},
\end{equation}
where $N_{\rm data}$ is the number of observed events in data, $\nObsMuMu$ for $\eta\to\gamma\mu^+\mu^-$ and $\nObsEE$ for $\eta\to\gamma e^+e^-$, $N_{\rm bkg}$ is the total number of background events as listed in Tables~\ref{tab_bkgmm} and~\ref{tab_bkgee}, $\epsilon$ is the efficiency estimated using signal MC samples, $\mathcal{B}(J/\psi\to\gamma\eta',\eta'\to\pi^+\pi^-\eta)$ is the BF quoted from the PDG. The measured BFs are $\mathcal{B}(\eta\to\gamma e^+e^-)=(\brEE\pm\brEEErrSta)\times 10^{-3}$ and $\mathcal{B}(\eta\to\gamma\mu^+\mu^-)=(\brMuMu\pm\brMuMuErrSta)\times 10^{-4}$.

\subsection{Systematic Uncertainties}\label{sec_sys}

The systematic uncertainties of the measured TFF of $\eta$ are from: (1) the photon detection, (2) the tracking and PID of charged particles, (3) the kinematic fit, (4) the background suppression criteria, and (5) the number of background events. The measured BFs suffer the same systematic uncertainties as mentioned above as well as the uncertainties in the number of $J/\psi$ events and the BFs quoted from the PDG.

The total number of $J/\psi$ events $(\nJpsi\pm\nJpsiErr)\times 10^{10}$ refers to Ref.~\cite{BESIII:2021cxx}, and its relative uncertainty is $\nJpsiErrR$\%. The quoted BFs are $\mathcal{B}(J/\psi\to\gamma\eta')=(5.25\pm0.07)\times10^{-3}$, and $\mathcal{B}(\eta'\to\pi^+\pi^-\eta)=(42.5\pm0.5)$\%, and their relative uncertainties are 1.34\% and 1.18\%, respectively. Systematic uncertainties from photon detection, tracking and PID of pions and electrons are evaluated using the control samples $e^+e^-\to\gamma\mu^+\mu^-$, $J/\psi\to\rho\pi$ and radiative Bhabha events at $\sqrt{s}=3.08$ GeV, respectively. For both data and MC simulation, the 2D efficiencies over the polar angle and energy (magnitude of momentum) of photons (charged particles) are given. For the TFF, we correct the signal MC efficiency by $\epsilon_{\rm data}/\epsilon_{\rm MC}$, where $\epsilon_{\rm data}$ and $\epsilon_{\rm MC}$ are the 2D efficiency distributions measured using control samples, and the resulting change of the TFF is taken as the corresponding uncertainty. For the BFs, the systematic uncertainties are weighted according to the angular and momentum distributions of data and MC simulation, and are taken as the corresponding uncertainties. For the slow momentum muons we assume the same uncertainties of the pions.

The uncertainty associated with the kinematic fit arises from the inconsistency of the $\chi^2$ distribution between data and MC simulation. The reconstructed energy and angle of the photons, the helix parameters of the charged tracks, and the related errors in MC simulations are corrected to make their distributions more consistent with data~\cite{BESIII:2012mpj}, thus obtaining better data-MC consistencies in the $\chi^2$ distributions. The corrected MC simulation is used for the nominal results. The difference in the efficiencies before and after this correction is taken as the uncertainty. 

The systematic uncertainty due to the photon conversion veto in the $\eta\to\gamma e^+e^-$ sample is estimated by changing the nominal criterion to 1.5 cm $< R_{xy} < 7.5$ cm, $\cos(\theta_{eg}) < 0.5$ and to 2.5 cm $< R_{xy} < 8.5$ cm, $\cos(\theta_{eg}) < 0.7$. The average change in the final results is assigned as the associated systematic uncertainty. To estimate the uncertainty of the number of background events, we vary each quoted BF by one standard deviation, and the resulting change is taken as the corresponding uncertainty.

The systematic uncertainties are summarized in Tables~\ref{tab_sysUFF} and~\ref{tab_sysUBr}.
	
\begin{table}[h]
\begin{center}
	\caption{Relative systematic uncertainties (in \%) for the $\Lambda^{-2}$ measurements.}\label{tab_sysUFF}
	\begin{tabular}{c|c|c}\hline\hline
		Source & $\eta\to\gamma\mu^+\mu^-$ & $\eta\to\gamma e^+e^-$ \\\hline
		Photon detection &$\sysUFFGamMuMu$ & $\sysUFFGamEE$\\
		Tracking & $\sysUFFTrkMuMu$ & $\sysUFFTrkEE$\\
		PID & $\sysUFFPIDMuMu$ & $\sysUFFPIDEE$\\
		Kinematic fit &$\sysUFFKmfitMuMu$ & $\sysUFFKmfitEE$\\
		$\gamma$ conversion veto & -- & $\sysUFFGamConvEE$\\
		Background & $\sysUFFBkgMuMu$ & $\sysUFFBkgEE$\\\hline
		Total &$\sysUFFTotalMuMu$ & $\sysUFFTotalEE$\\
		\hline\hline
	\end{tabular}
\end{center}
\end{table}
	
\begin{table}[h]
\begin{center}
	\caption{Relative systematic uncertainties (in \%) for the BF measurements.}\label{tab_sysUBr}
	\begin{tabular}{c|c|c}\hline\hline
		Source & $\eta\to\gamma\mu^+\mu^-$ & $\eta\to\gamma e^+e^-$ \\\hline
		Photon detection & $\sysUBFGamMuMu$ & $\sysUBFGamEE$\\
		Tracking & $\sysUBFTrkMuMu$ & $\sysUBFTrkEE$\\
		PID & $\sysUBFPIDMuMu$ & $\sysUBFPIDEE$\\
		Kinematic fit &$\sysUBFKmfitMuMu$ & $\sysUBFKmfitEE$\\
		Background  & $\sysUBFBkgMuMu$ & $\sysUBFBkgEE$\\
		$N_{J/\psi}$ & 0.4 &0.4\\
		$\gamma$ conversion veto & -- & $\sysUBFGamConvEE$\\
		$\mathcal{B}(J/\psi\to\gamma\eta')$ & 1.3 & 1.3\\
		$\mathcal{B}(\eta'\to\pi^+\pi^-\eta)$ & 1.2 & 1.2\\
		\hline
		Total &$\sysUBFTotalMuMu$ & $\sysUBFTotalEE$\\
		\hline\hline
	\end{tabular}
\end{center}
\end{table}

\section{Combination of Samples I and II}
\begin{figure}[htbp]
\begin{center}
	\subfigure[]{\includegraphics[width=.9\columnwidth]{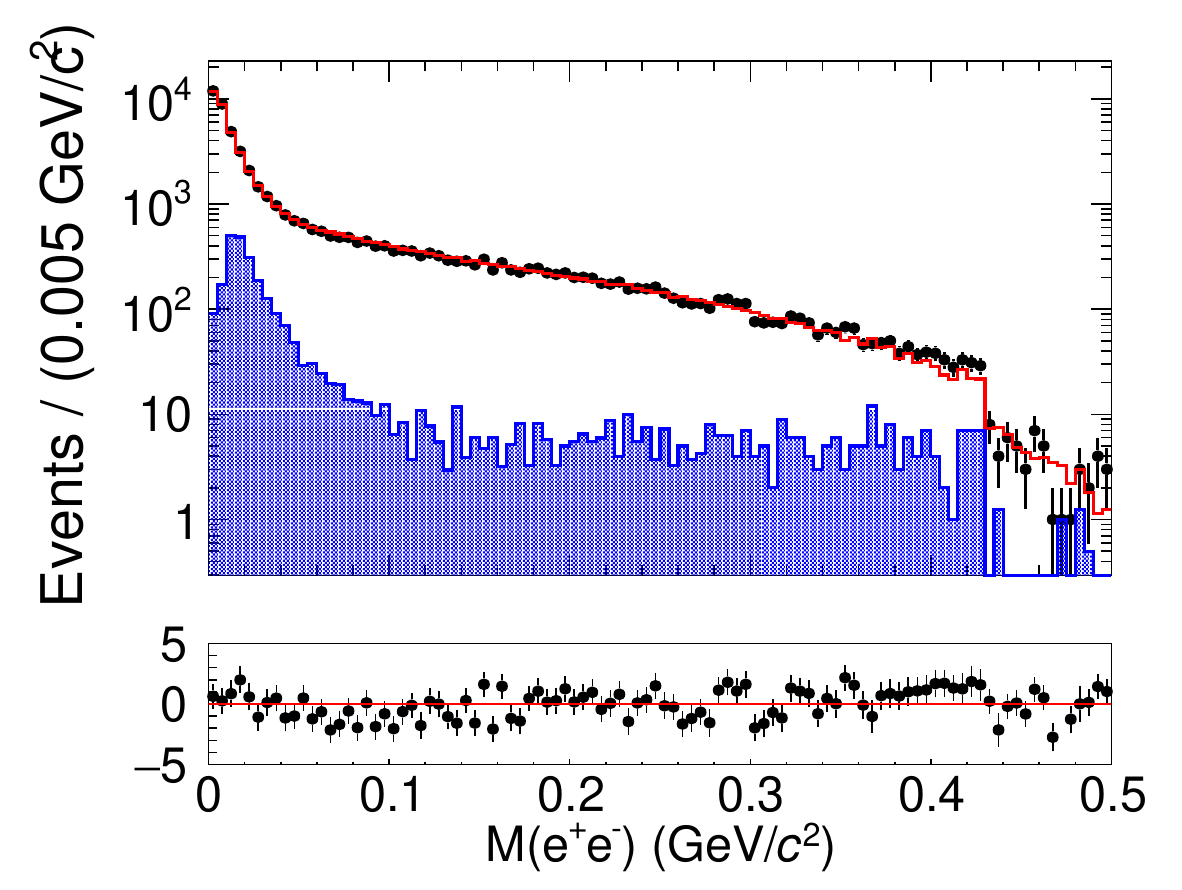}\label{fig_fitEEComb}}\\
	\subfigure[]{\includegraphics[width=.9\columnwidth]{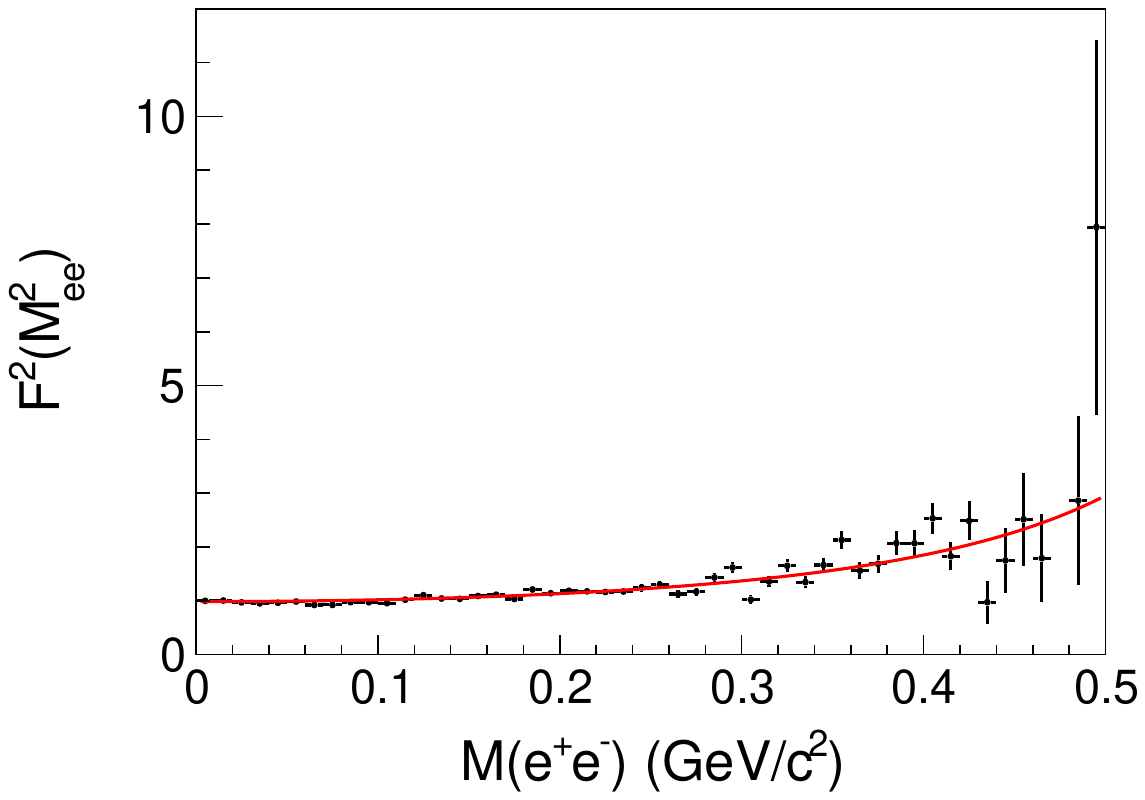}\label{fig_Combtff}}
	\caption{(a) Invariant mass distribution of $e^+e^-$ pairs. The black dots with error bars are from Samples I and II combined, the red solid line is the total fit result, including the signal and the background, the blue histogram is the background. (b) Distributions of $F^2$ as a function of $M(e^+e^-)$. The black dots with error bars are the ratios of the background-subtracted and efficiency-corrected data to the efficiency-corrected MC simulated using $F^2(q^2)\equiv 1$. The red line is the shape of the $F^2(q^2)$ function using the $\Lambda$ measured value.}
\end{center}
\end{figure}

To improve precision, an unbinned maximum likelihood fit is performed on the combined sample of Sample I and Sample II. Candidate events of Sample II are the same as the previous BESIII work~\cite{BESIII:2024pxo}. The likelihood is defined as:
\begin{equation}
	\mathcal{L}=\prod^N_{i=1}{\cal P}(\xi_i;\Lambda)\cdot\prod^M_{j=1}{\cal P}(\xi_j;\Lambda),
\end{equation}
where $N$ is the number of events in Sample I, and $M$ is the number of events in Sample II. By minimizing $S=-\ln\mathcal{L}_{\rm data}-(-\ln\mathcal{L}_{\rm bkg})$, the slope of the $\eta$ TFF is measured to be $\Lambda^{-2}(\eta\to\gamma e^+e^-)=\ffEEComb\pm\ffEEErrStaComb\pm\ffEEErrSysComb$ (GeV/$c^2$)$^{-2}$. The background of the two samples and the systematic uncertainties are estimated using the same methods as in Section~\ref{sec_ff} and Ref.~\cite{BESIII:2024pxo}. The systematic uncertainties are summarized in Table~\ref{tab_sysUFFComb}. The comparison of the fit results is shown in Fig.~\ref{fig_fitEEComb} and the TFF distributions of the combined data are shown in Fig.~\ref{fig_Combtff}.

\begin{table}[h]
\begin{center}
	\caption{Relative systematic uncertainties (in \%) for the combined $\Lambda^{-2}$ measurement.}\label{tab_sysUFFComb}
	\begin{tabular}{c|c}\hline\hline
		Source &  $\eta\to\gamma e^+e^-$ \\\hline
		Photon detection & 0.1\\
		Tracking &1.7\\
		PID &0.2\\
		Kinematic fit & $\sysUFFKmfitEE$\\
		Background  & 0.3\\
		$\gamma$ conversion veto & 0.2\\\hline
		Total & $\sysUFFTotalEEComb$\\
		\hline\hline
	\end{tabular}
\end{center}
\end{table}

The $\eta\to\gamma e^+e^-$ BFs measured in this and in the previous work~\cite{BESIII:2024pxo} are combined via:
\begin{equation}\label{eq_brComb}
	\bar{\mathcal{B}}=\frac{\Sigma_j(\mathcal{B}_j\cdot\Sigma_i\omega_{ij})}{\Sigma_i\Sigma_j\omega_{ij}}=\frac{\mathcal{B}_1\sigma_2^2+\mathcal{B}_2\sigma_1^2}{\sigma_1^2+\sigma_2^2+(\mathcal{B}_1-\mathcal{B}_2)^2\epsilon_f^2},
\end{equation}
where $i$ and $j$ are summed over all decay modes, $\mathcal{B}_j$ is the measured value given by the mode $j$, $\omega_{ij}$ is the element of the weight matrix $W=V^{-1}$, and $V$ is the covariance error matrix calculated as:
\begin{equation}\label{eq_V}
	V=
	\begin{pmatrix}
		\sigma_1^2+\epsilon_f^2\mathcal{B}_1^2 & \epsilon_f^2\mathcal{B}_1\mathcal{B}_2\\
		\epsilon_f^2\mathcal{B}_1\mathcal{B}_2 & \sigma_2^2+\epsilon_f^2\mathcal{B}_2^2
	\end{pmatrix},
\end{equation}
where $\sigma_i$ is the independent absolute uncertainty (includes the statistical uncertainty and all independent systematic uncertainties) in the mode $i$, and $\epsilon_f$ is the common relative systematic uncertainty between the two modes. The systematic uncertainties of the two modes have been measured in Section~\ref{sec_sys} and Ref.~\cite{BESIII:2024pxo}. They are summarized in Table~\ref{tab_sysUBrComb}.
The systematic uncertainty of $\bar{\mathcal{B}}$ is calculated as:
\begin{equation}\label{eq_brCombSysU}
	\sigma_{\bar{\mathcal{B}}}=\sqrt{\frac{1}{\Sigma_i\Sigma_j\omega_{ij}}}=\sqrt{\frac{\sigma_1^2\sigma_2^2+(\mathcal{B}_1^2\sigma_2^2+\mathcal{B}_2^2\sigma_1^2)\epsilon_f^2}{\sigma_1^2+\sigma_2^2+(\mathcal{B}_1-\mathcal{B}_2)^2\epsilon_f^2}}.
\end{equation}

Finally, the combined BF is $\mathcal{B}(\eta\to\gamma e^+e^-)=(\brEEComb\pm\brEEErrTotComb)\times 10^{-3}$.

\begin{table}[htbp]
\begin{center}
	\caption{Relative systematic uncertainties (in \%) for the combined $\mathcal{B}(\eta\to\gamma e^+e^-)$. The items marked with `*' are common uncertainties, and the other items are independent uncertainties.}\label{tab_sysUBrComb}
	\begin{tabular}{c|c|c}\hline\hline
		Source & Sample I& Sample II\\\hline
		Statistical & 0.04 & 0.05\\
		Photon detection* & $\sysUBFGamEE$ &0.5\\
		Tracking* &$\sysUBFTrkEE$ & 2.2\\
		PID* &$\sysUBFPIDEE$ & 0.9\\
		Kinematic fit & $\sysUBFKmfitEE$ & 0.3\\
		Background & $\sysUBFBkgEE$ & --\\
		$\gamma$ conversion veto & $\sysUBFGamConvEE$ & 0.9\\
		Number of peaking backgrounds & -- & 0.1\\
		Fit range and background shape & -- & 0.6\\
		Signal model& $\sysUBFModelEE$ & --\\
		$N_{J/\psi}$* & 0.4 & 0.4\\
		$\mathcal{B}(J/\psi\to\gamma\eta)$ & -- & 1.7\\
		$\mathcal{B}(J/\psi\to\gamma\eta')$ & 1.3 & --\\
		$\mathcal{B}(\eta'\to\pi^+\pi^-\eta)$ & 1.2 & --\\
		\hline
		Total &\multicolumn{2}{c}{$\sysUBFTotalEEComb$}\\
		\hline\hline
	\end{tabular}
\end{center}
\end{table}
	
\section{Search For the Dark Photon}
We search the dark photon $A'$ through its possible decay $A'\to e^+e^-$ by using the combined $\eta\to\gamma e^+e^-$ sample. The width of $A'$ is assumed to be zero, and the mass is scanned from 0.005 to 0.535 GeV/$c^2$ with a step length of 0.01 GeV/$c^2$. When scanning one assumed mass value of $A'$, a series of unbinned extended maximum likelihood fits on $M(e^+e^-)$ are performed. In each fit, the signal is described using the $\eta\rightarrow \gamma A',~A'\to e^+e^-$ MC shape and the number of signal events is free. The sizes and shapes of the backgrounds are described using the MC samples listed in Table~\ref{tab_bkgee} and the normalization of Sample I is fixed and free for Sample II. The corresponding likelihood can be obtained by fitting.

Through a series of fits, we obtain the likelihood distribution over a series of possible numbers of $A'$, and therefore the likelihood distribution over the possible BF of $\eta\to\gamma A',~A'\to e^+e^-$. As the most probable number of $A'$ signal events is around zero, the upper limit of $\mathcal{B}(\eta\to\gamma A',~A'\to e^+e^-)$ at 90\% confidence level is measured. We smear the likelihood distribution via,
\begin{equation}\label{eq_smearL}
	\mathcal{L}'(\mathcal{B})=\int^1_0 \mathcal{L}\left(\mathcal{B}\frac{r}{r_0}\right)\exp\left[-\frac{(r-r_0)^2}{2\delta_{r}^2} \right]dr,
\end{equation}
to take the systematic uncertainty, $\mathcal{B}(J/\psi\to\gamma\eta',~\eta'\to\pi^+\pi^-\eta)$, and $\mathcal{B}(J/\psi\to\gamma\eta)$ into consideration. Here, $\mathcal{B}$ stands for $\mathcal{B}(\eta\to\gamma A',~A'\to e^+e^-)$, $r_0\equiv\epsilon_0\cdot\mathcal{B}(J/\psi\to\gamma\eta',\eta'\to\pi^+\pi^-\eta)$ or $r_0\equiv\epsilon_0\cdot\mathcal{B}(J/\psi\to\gamma\eta)$, $\epsilon_0$ is the signal efficiency, and $\delta_{r}$ is the relative uncertainty of $r_0$. The uncertainty is estimated using the same way as described in Section~\ref{sec_sys} and Ref.~\cite{BESIII:2024pxo}. Then the upper limit, $\mathcal{B}_{\rm up}$, is determined as:
\begin{equation}\label{eq_upper}
	\int_0^{\mathcal{B}_{\rm up}}\mathcal{L'}(\mathcal{B})d\mathcal{B}/\int_0^{+\infty}\mathcal{L'}(\mathcal{B})d\mathcal{B}=0.9.
\end{equation}

\begin{figure}[htbp]
	\begin{center}
		\includegraphics[width=.9\columnwidth]{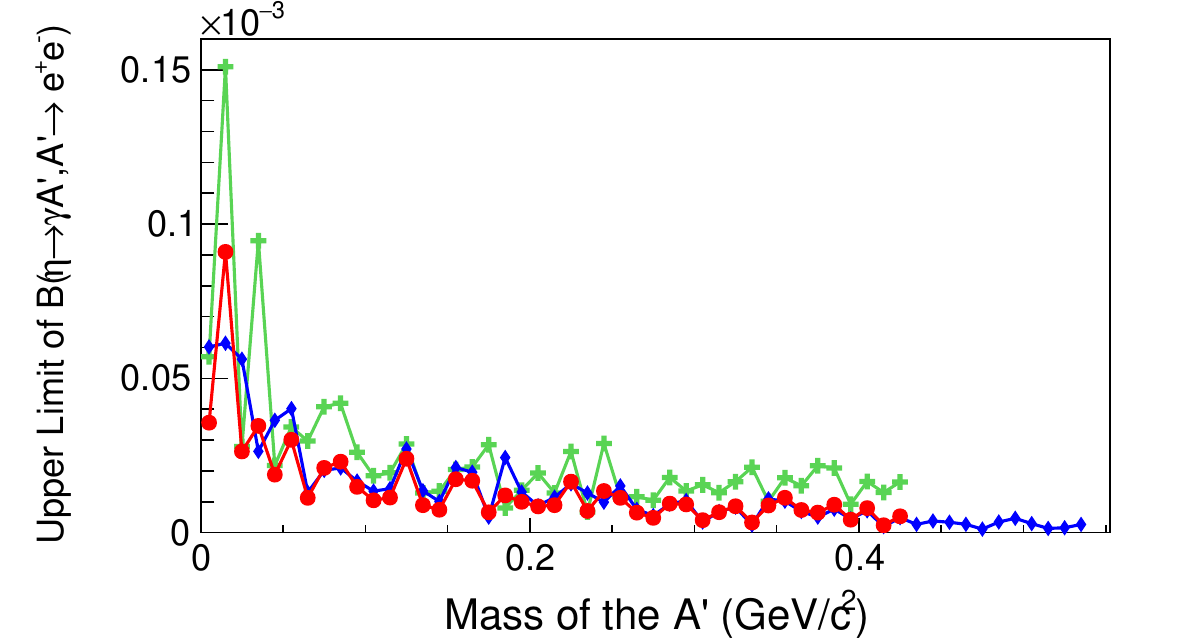}
		\caption{The upper limits of $\mathcal{B}(\eta\to\gamma A',~A'\to e^+e^-)$ of dark photons with different mass hypotheses. Blue diamonds, green crosses, and red circles are the upper limits measured with Sample I, Sample II, and the combined sample, respectively.}\label{fig_upperLimits}
	\end{center}
\end{figure}

\begin{figure}[htbp]
	\begin{center}
		\includegraphics[width=.9\columnwidth]{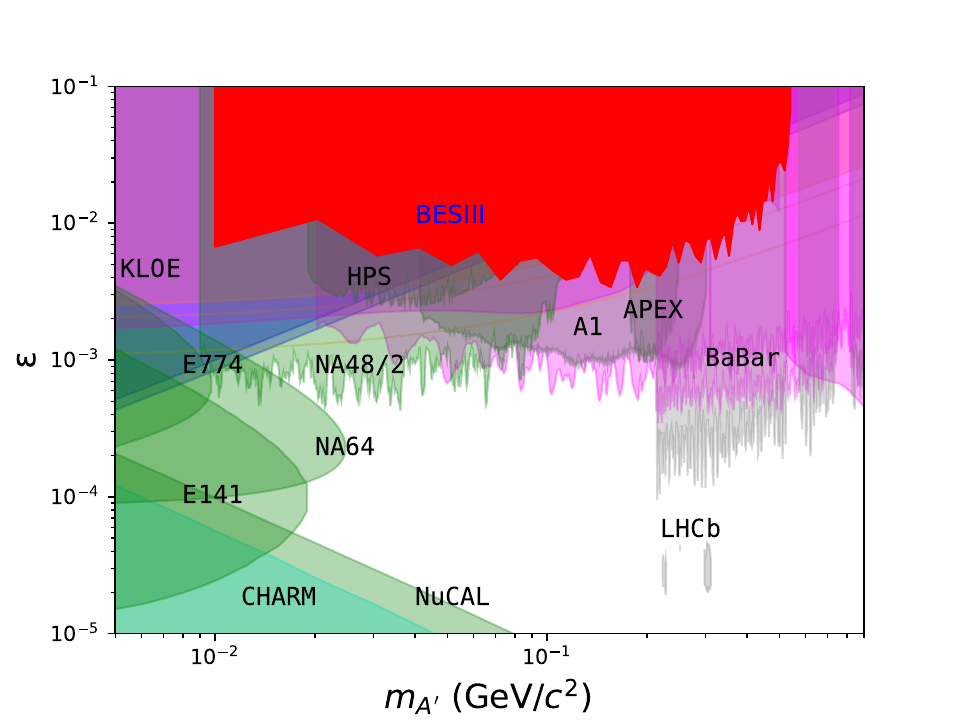}
		\caption{Exclusion limits at the 90\% confidence level on the mixing parameter $\epsilon$ as a function of the dark photon mass.}\label{fig_dplimits}
	\end{center}
\end{figure}

As the $r$ and $r_0$ of Samples I and II are different, we first measure the likelihood distributions using the two samples separately. Then the likelihoods are combined via:
\begin{equation}
	\mathcal{L}_{\rm comb}(\mathcal{B})=\mathcal{L}'_{\rm I}(\mathcal{B})\times\mathcal{L}'_{\rm II}(\mathcal{B}),
\end{equation}
where $\mathcal{L}'_{\rm I}(\mathcal{B})$ is the smeared likelihood for Sample I, and $\mathcal{L}'_{\rm II}(\mathcal{B})$ is for Sample II. The upper limits measured with each sample and the combined results are shown in Fig.~\ref{fig_upperLimits}. As the reconstruction efficiency of Sample II falls to zero in the region $M(e^+e^-) > 0.43$ GeV/$c^2$, only 43 $A'$ mass points are scanned for this sample. The comparison between our results and the other experiments is shown in Fig.~\ref{fig_dplimits}.

\section{Summary}
We propose a novel approach to measure the TFF of the $\eta$ meson using the $J/\psi\rightarrow\gamma\eta',~\eta'\to\pi^+\pi^-\eta$ decay. 
Based on 10 billion $J/\psi$ events collected by the BESIII detector, the analysis of the $\eta\to\gamma l^+l^-$ ($l=e,\mu$) decays is performed.

\begin{figure}[htbp]
	\centering
	\includegraphics[page=1,width=.85\columnwidth]{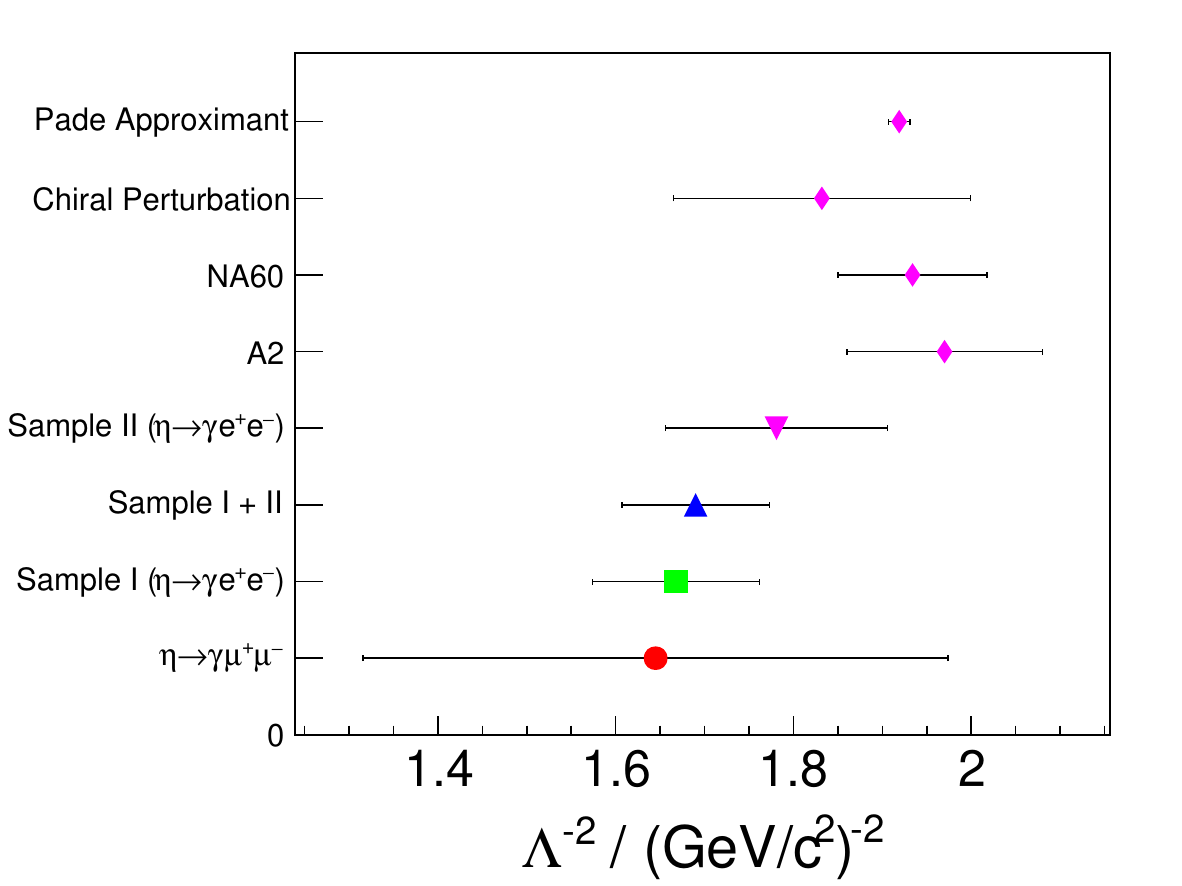}
	\caption{Comparison of the $\Lambda^{-2}$ values measured in this work and in previous works. From bottom to top, the results refer to this work, to previous works at BESIII~\cite{BESIII:2024pxo}, A2~\cite{NA60:2016nad}, and NA60~\cite{NA60:2016nad}, and to theoretical calculations~\cite{Bickert:2020kbn,Escribano:2015nra}.}
	\label{fig_compareFF}
\end{figure}

The BFs of $\eta\to\gamma l^+l^-$ are measured to be $\mathcal{B}(\eta\to\gamma e^+e^-)=(\brEE\pm\brEEErrSta\pm\brEEErrSys)\times 10^{-3}$ and $\mathcal{B}(\eta\to\gamma\mu^+\mu^-)=(\brMuMu\pm\brMuMuErrSta\pm\brMuMuErrSys)\times 10^{-4}$, which are both consistent with the previous measurements~\cite{ParticleDataGroup:2024cfk,Dzhelyadin:1980kh,Berghauser:2011zz}, but with improved precision. 
By investigating the $e^+e^-$ mass spectrum, the $\eta$ TFF is extracted to be 
$\Lambda^{-2}=\ffEE\pm\ffEEErrSta\pm\ffEEErrSys~({\rm GeV}/c^2)^{-2}$, 
which is in agreement with the previous BESIII result~\cite{BESIII:2024pxo}. It is slightly less than that from A2~\cite{Adlarson:2016hpp}, but in agreement within two standard deviations. For $\eta\to\gamma\mu^+\mu^-$, it is determined to be $\Lambda^{-2}=\ffMuMu\pm\ffMuMuErrSta\pm\ffMuMuErrSys$ (GeV/$c^2$)$^{-2}$, which is consistent with that from NA60~\cite{NA60:2016nad} within one standard deviation. Due to the limited statistics, the statistical uncertainty is dominant in the $\eta\rightarrow\gamma \mu^+\mu^-$ measurement. A comparison of the measured values of $\Lambda^{-2}$ is shown in Fig.~\ref{fig_compareFF}.

By means of a simultaneous analysis of the $J/\psi\to\gamma\eta',~\eta'\to\pi^+\pi^-\eta,~\eta\to\gamma e^+e^-$ decays and the $J/\psi\to\gamma\eta,~\eta\to\gamma e^+e^-$ decays performed in Ref.~\cite{BESIII:2024pxo}, the TFF and the BF are determined to be $\Lambda^{-2}(\eta\to\gamma e^+e^-)=\ffEEComb\pm\ffEEErrStaComb\pm\ffEEErrSysComb$ (GeV/$c^2$)$^{-2}$ and $\mathcal{B}(\eta\to\gamma e^+e^-)=(\brEEComb\pm\brEEErrTotComb)\times 10^{-3}$, respectively. In addition, we search for the dark photon, denoted by $A'$, using the combined events. As no signal is observed, the upper limits of $\mathcal{B}(\eta\to\gamma A',~A'\to e^+e^-)$ at 90\% confidence level for $A'$ with different masses are given, as shown in Fig.~\ref{fig_upperLimits} and~\ref{fig_dplimits}.

\section*{ACKNOWLEDGMENTS}
The BESIII Collaboration thanks the staff of BEPCII and the IHEP computing center for their strong support. This work is supported in part by National Key R\&D Program of China under Contracts Nos. 2023YFA1606000; National Natural Science Foundation of China (NSFC) under Contracts Nos. 11635010, 11735014, 11935015, 11935016, 11935018, 12025502, 12035009, 12035013, 12061131003, 12192260, 12192261, 12192262, 12192263, 12192264, 12192265, 12221005, 12225509, 12235017, 12361141819; the Chinese Academy of Sciences (CAS) Large-Scale Scientific Facility Program; the CAS Center for Excellence in Particle Physics (CCEPP); Joint Large-Scale Scientific Facility Funds of the NSFC and CAS under Contract No. U1832207; CAS under Contract No. YSBR-101; 100 Talents Program of CAS; The Institute of Nuclear and Particle Physics (INPAC) and Shanghai Key Laboratory for Particle Physics and Cosmology; Agencia Nacional de Investigación y Desarrollo de Chile (ANID), Chile under Contract No. ANID PIA/APOYO AFB230003; German Research Foundation DFG under Contract No. FOR5327; Istituto Nazionale di Fisica Nucleare, Italy; Knut and Alice Wallenberg Foundation under Contracts Nos. 2021.0174, 2021.0299; Ministry of Development of Turkey under Contract No. DPT2006K-120470; National Research Foundation of Korea under Contract No. NRF-2022R1A2C1092335; National Science and Technology fund of Mongolia; National Science Research and Innovation Fund (NSRF) via the Program Management Unit for Human Resources \& Institutional Development, Research and Innovation of Thailand under Contract No. B50G670107; Polish National Science Centre under Contract No. 2019/35/O/ST2/02907; Swedish Research Council under Contract No. 2019.04595; The Swedish Foundation for International Cooperation in Research and Higher Education under Contract No. CH2018-7756; U. S. Department of Energy under Contract No. DE-FG02-05ER41374; This paper is also supported by Guangdong Basic and Applied Basic Research Foundation 2024A1515012416

\bibliographystyle{apsrev4-1}
\bibliography{main}

\begin{thebibliography}{37}%
\makeatletter
\providecommand \@ifxundefined [1]{%
 \@ifx{#1\undefined}
}%
\providecommand \@ifnum [1]{%
 \ifnum #1\expandafter \@firstoftwo
 \else \expandafter \@secondoftwo
 \fi
}%
\providecommand \@ifx [1]{%
 \ifx #1\expandafter \@firstoftwo
 \else \expandafter \@secondoftwo
 \fi
}%
\providecommand \natexlab [1]{#1}%
\providecommand \enquote  [1]{``#1''}%
\providecommand \bibnamefont  [1]{#1}%
\providecommand \bibfnamefont [1]{#1}%
\providecommand \citenamefont [1]{#1}%
\providecommand \href@noop [0]{\@secondoftwo}%
\providecommand \href [0]{\begingroup \@sanitize@url \@href}%
\providecommand \@href[1]{\@@startlink{#1}\@@href}%
\providecommand \@@href[1]{\endgroup#1\@@endlink}%
\providecommand \@sanitize@url [0]{\catcode `\\12\catcode `\$12\catcode `\&12\catcode `\#12\catcode `\^12\catcode `\_12\catcode `\%12\relax}%
\providecommand \@@startlink[1]{}%
\providecommand \@@endlink[0]{}%
\providecommand \url  [0]{\begingroup\@sanitize@url \@url }%
\providecommand \@url [1]{\endgroup\@href {#1}{\urlprefix }}%
\providecommand \urlprefix  [0]{URL }%
\providecommand \Eprint [0]{\href }%
\providecommand \doibase [0]{http://dx.doi.org/}%
\providecommand \selectlanguage [0]{\@gobble}%
\providecommand \bibinfo  [0]{\@secondoftwo}%
\providecommand \bibfield  [0]{\@secondoftwo}%
\providecommand \translation [1]{[#1]}%
\providecommand \BibitemOpen [0]{}%
\providecommand \bibitemStop [0]{}%
\providecommand \bibitemNoStop [0]{.\EOS\space}%
\providecommand \EOS [0]{\spacefactor3000\relax}%
\providecommand \BibitemShut  [1]{\csname bibitem#1\endcsname}%
\let\auto@bib@innerbib\@empty
\bibitem [{\citenamefont {Aoyama}\ \emph {et~al.}(2020)\citenamefont {Aoyama} \emph {et~al.}}]{Aoyama:2020ynm}%
  \BibitemOpen
  \bibfield  {author} {\bibinfo {author} {\bibfnamefont {T.}~\bibnamefont {Aoyama}} \emph {et~al.},\ }\href {\doibase 10.1016/j.physrep.2020.07.006} {\bibfield  {journal} {\bibinfo  {journal} {Phys. Rept.}\ }\textbf {\bibinfo {volume} {887}},\ \bibinfo {pages} {1} (\bibinfo {year} {2020})}\BibitemShut {NoStop}%
\bibitem [{\citenamefont {Landsberg}(1985)}]{Landsberg:1985gaz}%
  \BibitemOpen
  \bibfield  {author} {\bibinfo {author} {\bibfnamefont {L.~G.}\ \bibnamefont {Landsberg}},\ }\href {\doibase 10.1016/0370-1573(85)90129-2} {\bibfield  {journal} {\bibinfo  {journal} {Phys. Rept.}\ }\textbf {\bibinfo {volume} {128}},\ \bibinfo {pages} {301} (\bibinfo {year} {1985})}\BibitemShut {NoStop}%
\bibitem [{\citenamefont {Adlarson}\ \emph {et~al.}(2017)\citenamefont {Adlarson} \emph {et~al.}}]{Adlarson:2016hpp}%
  \BibitemOpen
  \bibfield  {author} {\bibinfo {author} {\bibfnamefont {P.}~\bibnamefont {Adlarson}} \emph {et~al.},\ }\href {\doibase 10.1103/PhysRevC.95.035208} {\bibfield  {journal} {\bibinfo  {journal} {Phys. Rev. C}\ }\textbf {\bibinfo {volume} {95}},\ \bibinfo {pages} {035208} (\bibinfo {year} {2017})}\BibitemShut {NoStop}%
\bibitem [{\citenamefont {Arnaldi}\ \emph {et~al.}(2016)\citenamefont {Arnaldi} \emph {et~al.}}]{NA60:2016nad}%
  \BibitemOpen
  \bibfield  {author} {\bibinfo {author} {\bibfnamefont {R.}~\bibnamefont {Arnaldi}} \emph {et~al.} (\bibinfo {collaboration} {NA60 Collaboration}),\ }\href {\doibase 10.1016/j.physletb.2016.04.013} {\bibfield  {journal} {\bibinfo  {journal} {Phys. Lett. B}\ }\textbf {\bibinfo {volume} {757}},\ \bibinfo {pages} {437} (\bibinfo {year} {2016})}\BibitemShut {NoStop}%
\bibitem [{\citenamefont {Ablikim}\ \emph {et~al.}(2024{\natexlab{a}})\citenamefont {Ablikim} \emph {et~al.}}]{BESIII:2024pxo}%
  \BibitemOpen
  \bibfield  {author} {\bibinfo {author} {\bibfnamefont {M.}~\bibnamefont {Ablikim}} \emph {et~al.} (\bibinfo {collaboration} {BESIII Collaboration}),\ }\href {\doibase 10.1103/PhysRevD.109.072001} {\bibfield  {journal} {\bibinfo  {journal} {Phys. Rev. D}\ }\textbf {\bibinfo {volume} {109}},\ \bibinfo {pages} {072001} (\bibinfo {year} {2024}{\natexlab{a}})}\BibitemShut {NoStop}%
\bibitem [{\citenamefont {Ablikim}\ \emph {et~al.}(2022)\citenamefont {Ablikim} \emph {et~al.}}]{BESIII:2021cxx}%
  \BibitemOpen
  \bibfield  {author} {\bibinfo {author} {\bibfnamefont {M.}~\bibnamefont {Ablikim}} \emph {et~al.} (\bibinfo {collaboration} {BESIII Collaboration}),\ }\href {\doibase 10.1088/1674-1137/ac5c2e} {\bibfield  {journal} {\bibinfo  {journal} {Chin. Phys. C}\ }\textbf {\bibinfo {volume} {46}},\ \bibinfo {pages} {074001} (\bibinfo {year} {2022})}\BibitemShut {NoStop}%
\bibitem [{\citenamefont {Gan}\ \emph {et~al.}(2022)\citenamefont {Gan}, \citenamefont {Kubis}, \citenamefont {Passemar},\ and\ \citenamefont {Tulin}}]{Gan:2020aco}%
  \BibitemOpen
  \bibfield  {author} {\bibinfo {author} {\bibfnamefont {L.}~\bibnamefont {Gan}}, \bibinfo {author} {\bibfnamefont {B.}~\bibnamefont {Kubis}}, \bibinfo {author} {\bibfnamefont {E.}~\bibnamefont {Passemar}}, \ and\ \bibinfo {author} {\bibfnamefont {S.}~\bibnamefont {Tulin}},\ }\href {\doibase 10.1016/j.physrep.2021.11.001} {\bibfield  {journal} {\bibinfo  {journal} {Phys. Rept.}\ }\textbf {\bibinfo {volume} {945}},\ \bibinfo {pages} {1} (\bibinfo {year} {2022})}\BibitemShut {NoStop}%
\bibitem [{\citenamefont {Ablikim}\ \emph {et~al.}(2010)\citenamefont {Ablikim} \emph {et~al.}}]{BESIII:2009fln}%
  \BibitemOpen
  \bibfield  {author} {\bibinfo {author} {\bibfnamefont {M.}~\bibnamefont {Ablikim}} \emph {et~al.} (\bibinfo {collaboration} {BESIII Collaboration}),\ }\href {\doibase 10.1016/j.nima.2009.12.050} {\bibfield  {journal} {\bibinfo  {journal} {Nucl. Instrum. Meth. A}\ }\textbf {\bibinfo {volume} {614}},\ \bibinfo {pages} {345} (\bibinfo {year} {2010})}\BibitemShut {NoStop}%
\bibitem [{\citenamefont {Yu}\ \emph {et~al.}(2016)\citenamefont {Yu} \emph {et~al.}}]{Yu:2016cof}%
  \BibitemOpen
  \bibfield  {author} {\bibinfo {author} {\bibfnamefont {C.}~\bibnamefont {Yu}} \emph {et~al.},\ }in\ \href {\doibase 10.18429/JACoW-IPAC2016-TUYA01} {\emph {\bibinfo {booktitle} {{7th International Particle Accelerator Conference}}}}\ (\bibinfo {year} {2016})\ p.\ \bibinfo {pages} {TUYA01}\BibitemShut {NoStop}%
\bibitem [{\citenamefont {Ablikim}\ \emph {et~al.}(2020)\citenamefont {Ablikim} \emph {et~al.}}]{BESIII:2020nme}%
  \BibitemOpen
  \bibfield  {author} {\bibinfo {author} {\bibfnamefont {M.}~\bibnamefont {Ablikim}} \emph {et~al.} (\bibinfo {collaboration} {BESIII Collaboration}),\ }\href {\doibase 10.1088/1674-1137/44/4/040001} {\bibfield  {journal} {\bibinfo  {journal} {Chin. Phys. C}\ }\textbf {\bibinfo {volume} {44}},\ \bibinfo {pages} {040001} (\bibinfo {year} {2020})}\BibitemShut {NoStop}%
\bibitem [{\citenamefont {Jiao}\ \emph {et~al.}(2020)\citenamefont {Jiao}, \citenamefont {Chen}, \citenamefont {He}, \citenamefont {Li}, \citenamefont {Li}, \citenamefont {Qin}, \citenamefont {Qu}, \citenamefont {Wan}, \citenamefont {Wang},\ and\ \citenamefont {Xu}}]{Jiao:2020dqs}%
  \BibitemOpen
  \bibfield  {author} {\bibinfo {author} {\bibfnamefont {Y.}~\bibnamefont {Jiao}}, \bibinfo {author} {\bibfnamefont {F.}~\bibnamefont {Chen}}, \bibinfo {author} {\bibfnamefont {P.}~\bibnamefont {He}}, \bibinfo {author} {\bibfnamefont {C.}~\bibnamefont {Li}}, \bibinfo {author} {\bibfnamefont {J.}~\bibnamefont {Li}}, \bibinfo {author} {\bibfnamefont {Q.}~\bibnamefont {Qin}}, \bibinfo {author} {\bibfnamefont {H.}~\bibnamefont {Qu}}, \bibinfo {author} {\bibfnamefont {J.}~\bibnamefont {Wan}}, \bibinfo {author} {\bibfnamefont {J.}~\bibnamefont {Wang}}, \ and\ \bibinfo {author} {\bibfnamefont {G.}~\bibnamefont {Xu}},\ }\href {\doibase 10.1007/s41605-020-00189-7} {\bibfield  {journal} {\bibinfo  {journal} {Rad. Det. Tech. Meth.}\ }\textbf {\bibinfo {volume} {4}},\ \bibinfo {pages} {415} (\bibinfo {year} {2020})}\BibitemShut {NoStop}%
\bibitem [{\citenamefont {Zhang}\ \emph {et~al.}(2022)\citenamefont {Zhang} \emph {et~al.}}]{Zhang:2022bdc}%
  \BibitemOpen
  \bibfield  {author} {\bibinfo {author} {\bibfnamefont {J.-W.}\ \bibnamefont {Zhang}} \emph {et~al.},\ }\href {\doibase 10.1007/s41605-022-00331-7} {\bibfield  {journal} {\bibinfo  {journal} {Rad. Det. Tech. Meth.}\ }\textbf {\bibinfo {volume} {6}},\ \bibinfo {pages} {289} (\bibinfo {year} {2022})}\BibitemShut {NoStop}%
\bibitem [{\citenamefont {Cao}\ \emph {et~al.}(2020)\citenamefont {Cao} \emph {et~al.}}]{Cao:2020ibk}%
  \BibitemOpen
  \bibfield  {author} {\bibinfo {author} {\bibfnamefont {P.}~\bibnamefont {Cao}} \emph {et~al.},\ }\href {\doibase 10.1016/j.nima.2019.163053} {\bibfield  {journal} {\bibinfo  {journal} {Nucl. Instrum. Meth. A}\ }\textbf {\bibinfo {volume} {953}},\ \bibinfo {pages} {163053} (\bibinfo {year} {2020})}\BibitemShut {NoStop}%
\bibitem [{\citenamefont {Agostinelli}\ \emph {et~al.}(2003)\citenamefont {Agostinelli} \emph {et~al.}}]{GEANT4:2002zbu}%
  \BibitemOpen
  \bibfield  {author} {\bibinfo {author} {\bibfnamefont {S.}~\bibnamefont {Agostinelli}} \emph {et~al.} (\bibinfo {collaboration} {GEANT4 Collaboration}),\ }\href {\doibase 10.1016/S0168-9002(03)01368-8} {\bibfield  {journal} {\bibinfo  {journal} {Nucl. Instrum. Meth. A}\ }\textbf {\bibinfo {volume} {506}},\ \bibinfo {pages} {250} (\bibinfo {year} {2003})}\BibitemShut {NoStop}%
\bibitem [{\citenamefont {Yan}\ \emph {et~al.}(2006)\citenamefont {Yan}, \citenamefont {Fu}, \citenamefont {Dong}, \citenamefont {Miao}, \citenamefont {Huai-Min}, \citenamefont {Jun}, \citenamefont {Yu}, \citenamefont {Yun}, \citenamefont {Ye}, \citenamefont {Rui}, \citenamefont {Jie}, \citenamefont {Mei},\ and\ \citenamefont {Xiang}}]{MCPackage}%
  \BibitemOpen
  \bibfield  {author} {\bibinfo {author} {\bibfnamefont {D.~Z.}\ \bibnamefont {Yan}}, \bibinfo {author} {\bibfnamefont {C.~G.}\ \bibnamefont {Fu}}, \bibinfo {author} {\bibfnamefont {F.~C.}\ \bibnamefont {Dong}}, \bibinfo {author} {\bibfnamefont {H.}~\bibnamefont {Miao}}, \bibinfo {author} {\bibfnamefont {L.}~\bibnamefont {Huai-Min}}, \bibinfo {author} {\bibfnamefont {M.~Y.}\ \bibnamefont {Jun}}, \bibinfo {author} {\bibfnamefont {X.}~\bibnamefont {Yu}}, \bibinfo {author} {\bibfnamefont {Y.~Z.}\ \bibnamefont {Yun}}, \bibinfo {author} {\bibfnamefont {Y.}~\bibnamefont {Ye}}, \bibinfo {author} {\bibfnamefont {T.}~\bibnamefont {Rui}}, \bibinfo {author} {\bibfnamefont {L.~Y.}\ \bibnamefont {Jie}}, \bibinfo {author} {\bibfnamefont {M.~Q.}\ \bibnamefont {Mei}}, \ and\ \bibinfo {author} {\bibfnamefont {M.}~\bibnamefont {Xiang}},\ }\href {http://hepnp.ihep.ac.cn/en/article/id/283d17c0-e8fa-4ad7-bfe3-92095466def1} {\bibfield  {journal} {\bibinfo  {journal} {Chinese Physics C}\ }\textbf {\bibinfo {volume} {30}},\ \bibinfo
  {pages} {371} (\bibinfo {year} {2006})}\BibitemShut {NoStop}%
\bibitem [{\citenamefont {Yun}\ \emph {et~al.}(2008)\citenamefont {Yun}, \citenamefont {Tie},\ and\ \citenamefont {Jun}}]{You_2008}%
  \BibitemOpen
  \bibfield  {author} {\bibinfo {author} {\bibfnamefont {Y.~Z.}\ \bibnamefont {Yun}}, \bibinfo {author} {\bibfnamefont {L.~Y.}\ \bibnamefont {Tie}}, \ and\ \bibinfo {author} {\bibfnamefont {M.~Y.}\ \bibnamefont {Jun}},\ }\href {\doibase 10.1088/1674-1137/32/7/012} {\bibfield  {journal} {\bibinfo  {journal} {Chinese Physics C}\ }\textbf {\bibinfo {volume} {32}},\ \bibinfo {pages} {572} (\bibinfo {year} {2008})}\BibitemShut {NoStop}%
\bibitem [{\citenamefont {Liang}\ \emph {et~al.}(2009)\citenamefont {Liang} \emph {et~al.}}]{Liang:2009zzb}%
  \BibitemOpen
  \bibfield  {author} {\bibinfo {author} {\bibfnamefont {Y.~T.}\ \bibnamefont {Liang}} \emph {et~al.},\ }\href {\doibase 10.1016/j.nima.2009.02.036} {\bibfield  {journal} {\bibinfo  {journal} {Nucl. Instrum. Meth. A}\ }\textbf {\bibinfo {volume} {603}},\ \bibinfo {pages} {325} (\bibinfo {year} {2009})}\BibitemShut {NoStop}%
\bibitem [{\citenamefont {Jadach}\ \emph {et~al.}(2001)\citenamefont {Jadach}, \citenamefont {Ward},\ and\ \citenamefont {Was}}]{Jadach:2000ir}%
  \BibitemOpen
  \bibfield  {author} {\bibinfo {author} {\bibfnamefont {S.}~\bibnamefont {Jadach}}, \bibinfo {author} {\bibfnamefont {B.~F.~L.}\ \bibnamefont {Ward}}, \ and\ \bibinfo {author} {\bibfnamefont {Z.}~\bibnamefont {Was}},\ }\href {\doibase 10.1103/PhysRevD.63.113009} {\bibfield  {journal} {\bibinfo  {journal} {Phys. Rev. D}\ }\textbf {\bibinfo {volume} {63}},\ \bibinfo {pages} {113009} (\bibinfo {year} {2001})}\BibitemShut {NoStop}%
\bibitem [{\citenamefont {Ping}(2008)}]{Ping:2008zz}%
  \BibitemOpen
  \bibfield  {author} {\bibinfo {author} {\bibfnamefont {R.~G.}\ \bibnamefont {Ping}},\ }\href {\doibase 10.1088/1674-1137/32/8/001} {\bibfield  {journal} {\bibinfo  {journal} {Chin. Phys. C}\ }\textbf {\bibinfo {volume} {32}},\ \bibinfo {pages} {599} (\bibinfo {year} {2008})}\BibitemShut {NoStop}%
\bibitem [{\citenamefont {Navas}\ \emph {et~al.}(2024)\citenamefont {Navas} \emph {et~al.}}]{ParticleDataGroup:2024cfk}%
  \BibitemOpen
  \bibfield  {author} {\bibinfo {author} {\bibfnamefont {S.}~\bibnamefont {Navas}} \emph {et~al.} (\bibinfo {collaboration} {Particle Data Group}),\ }\href {\doibase 10.1103/PhysRevD.110.030001} {\bibfield  {journal} {\bibinfo  {journal} {Phys. Rev. D}\ }\textbf {\bibinfo {volume} {110}},\ \bibinfo {pages} {030001} (\bibinfo {year} {2024})}\BibitemShut {NoStop}%
\bibitem [{\citenamefont {Chen}\ \emph {et~al.}(2000)\citenamefont {Chen}, \citenamefont {Huang}, \citenamefont {Qi}, \citenamefont {Zhang},\ and\ \citenamefont {Zhu}}]{Chen:2000tv}%
  \BibitemOpen
  \bibfield  {author} {\bibinfo {author} {\bibfnamefont {J.~C.}\ \bibnamefont {Chen}}, \bibinfo {author} {\bibfnamefont {G.~S.}\ \bibnamefont {Huang}}, \bibinfo {author} {\bibfnamefont {X.~R.}\ \bibnamefont {Qi}}, \bibinfo {author} {\bibfnamefont {D.~H.}\ \bibnamefont {Zhang}}, \ and\ \bibinfo {author} {\bibfnamefont {Y.~S.}\ \bibnamefont {Zhu}},\ }\href {\doibase 10.1103/PhysRevD.62.034003} {\bibfield  {journal} {\bibinfo  {journal} {Phys. Rev. D}\ }\textbf {\bibinfo {volume} {62}},\ \bibinfo {pages} {034003} (\bibinfo {year} {2000})}\BibitemShut {NoStop}%
\bibitem [{\citenamefont {Yang}\ \emph {et~al.}(2014)\citenamefont {Yang}, \citenamefont {Ping},\ and\ \citenamefont {Chen}}]{Yang:2014vra}%
  \BibitemOpen
  \bibfield  {author} {\bibinfo {author} {\bibfnamefont {R.~L.}\ \bibnamefont {Yang}}, \bibinfo {author} {\bibfnamefont {R.~G.}\ \bibnamefont {Ping}}, \ and\ \bibinfo {author} {\bibfnamefont {H.}~\bibnamefont {Chen}},\ }\href {\doibase 10.1088/0256-307X/31/6/061301} {\bibfield  {journal} {\bibinfo  {journal} {Chin. Phys. Lett.}\ }\textbf {\bibinfo {volume} {31}},\ \bibinfo {pages} {061301} (\bibinfo {year} {2014})}\BibitemShut {NoStop}%
\bibitem [{\citenamefont {Richter-Was}(1993)}]{Richter-Was:1992hxq}%
  \BibitemOpen
  \bibfield  {author} {\bibinfo {author} {\bibfnamefont {E.}~\bibnamefont {Richter-Was}},\ }\href {\doibase 10.1016/0370-2693(93)90062-M} {\bibfield  {journal} {\bibinfo  {journal} {Phys. Lett. B}\ }\textbf {\bibinfo {volume} {303}},\ \bibinfo {pages} {163} (\bibinfo {year} {1993})}\BibitemShut {NoStop}%
\bibitem [{\citenamefont {Morisita}\ \emph {et~al.}(1991)\citenamefont {Morisita}, \citenamefont {Kitamura},\ and\ \citenamefont {Teshima}}]{Morisita:1990cg}%
  \BibitemOpen
  \bibfield  {author} {\bibinfo {author} {\bibfnamefont {N.}~\bibnamefont {Morisita}}, \bibinfo {author} {\bibfnamefont {I.}~\bibnamefont {Kitamura}}, \ and\ \bibinfo {author} {\bibfnamefont {T.}~\bibnamefont {Teshima}},\ }\href {\doibase 10.1103/PhysRevD.44.175} {\bibfield  {journal} {\bibinfo  {journal} {Phys. Rev. D}\ }\textbf {\bibinfo {volume} {44}},\ \bibinfo {pages} {175} (\bibinfo {year} {1991})}\BibitemShut {NoStop}%
\bibitem [{\citenamefont {Ablikim}\ \emph {et~al.}(2018{\natexlab{a}})\citenamefont {Ablikim} \emph {et~al.}}]{BESIII:2017djm}%
  \BibitemOpen
  \bibfield  {author} {\bibinfo {author} {\bibfnamefont {M.}~\bibnamefont {Ablikim}} \emph {et~al.} (\bibinfo {collaboration} {BESIII Collaboration}),\ }\href {\doibase 10.1103/PhysRevD.97.012003} {\bibfield  {journal} {\bibinfo  {journal} {Phys. Rev. D}\ }\textbf {\bibinfo {volume} {97}},\ \bibinfo {pages} {012003} (\bibinfo {year} {2018}{\natexlab{a}})}\BibitemShut {NoStop}%
\bibitem [{\citenamefont {Dzhelyadin}\ \emph {et~al.}(1980)\citenamefont {Dzhelyadin} \emph {et~al.}}]{Dzhelyadin:1980kh}%
  \BibitemOpen
  \bibfield  {author} {\bibinfo {author} {\bibfnamefont {R.~I.}\ \bibnamefont {Dzhelyadin}} \emph {et~al.},\ }\href {\doibase 10.1016/0370-2693(80)90937-5} {\bibfield  {journal} {\bibinfo  {journal} {Phys. Lett. B}\ }\textbf {\bibinfo {volume} {94}},\ \bibinfo {pages} {548} (\bibinfo {year} {1980})}\BibitemShut {NoStop}%
\bibitem [{\citenamefont {Berghauser}\ \emph {et~al.}(2011)\citenamefont {Berghauser} \emph {et~al.}}]{Berghauser:2011zz}%
  \BibitemOpen
  \bibfield  {author} {\bibinfo {author} {\bibfnamefont {H.}~\bibnamefont {Berghauser}} \emph {et~al.},\ }\href {\doibase 10.1016/j.physletb.2011.06.069} {\bibfield  {journal} {\bibinfo  {journal} {Phys. Lett. B}\ }\textbf {\bibinfo {volume} {701}},\ \bibinfo {pages} {562} (\bibinfo {year} {2011})}\BibitemShut {NoStop}%
\bibitem [{\citenamefont {Ablikim}\ \emph {et~al.}(2018{\natexlab{b}})\citenamefont {Ablikim} \emph {et~al.}}]{BESIII:2017kyd}%
  \BibitemOpen
  \bibfield  {author} {\bibinfo {author} {\bibfnamefont {M.}~\bibnamefont {Ablikim}} \emph {et~al.} (\bibinfo {collaboration} {BESIII Collaboration}),\ }\href {\doibase 10.1103/PhysRevLett.120.242003} {\bibfield  {journal} {\bibinfo  {journal} {Phys. Rev. Lett.}\ }\textbf {\bibinfo {volume} {120}},\ \bibinfo {pages} {242003} (\bibinfo {year} {2018}{\natexlab{b}})}\BibitemShut {NoStop}%
\bibitem [{\citenamefont {Zhang}\ \emph {et~al.}(2012)\citenamefont {Zhang}, \citenamefont {Qin},\ and\ \citenamefont {Fang}}]{Zhang:2012gt}%
  \BibitemOpen
  \bibfield  {author} {\bibinfo {author} {\bibfnamefont {Z.~Y.}\ \bibnamefont {Zhang}}, \bibinfo {author} {\bibfnamefont {L.~Q.}\ \bibnamefont {Qin}}, \ and\ \bibinfo {author} {\bibfnamefont {S.~S.}\ \bibnamefont {Fang}},\ }\href {\doibase 10.1088/1674-1137/36/10/002} {\bibfield  {journal} {\bibinfo  {journal} {Chin. Phys. C}\ }\textbf {\bibinfo {volume} {36}},\ \bibinfo {pages} {926} (\bibinfo {year} {2012})}\BibitemShut {NoStop}%
\bibitem [{\citenamefont {Ablikim}\ \emph {et~al.}(2024{\natexlab{b}})\citenamefont {Ablikim} \emph {et~al.}}]{BESIII:2024awu}%
  \BibitemOpen
  \bibfield  {author} {\bibinfo {author} {\bibfnamefont {M.}~\bibnamefont {Ablikim}} \emph {et~al.} (\bibinfo {collaboration} {BESIII Collaboration}),\ }\href {\doibase 10.1007/JHEP07(2024)135} {\bibfield  {journal} {\bibinfo  {journal} {JHEP}\ }\textbf {\bibinfo {volume} {07}},\ \bibinfo {pages} {135} (\bibinfo {year} {2024}{\natexlab{b}})}\BibitemShut {NoStop}%
\bibitem [{\citenamefont {Xu}\ and\ \citenamefont {He}(2012)}]{Xu:2012xq}%
  \BibitemOpen
  \bibfield  {author} {\bibinfo {author} {\bibfnamefont {Z.~R.}\ \bibnamefont {Xu}}\ and\ \bibinfo {author} {\bibfnamefont {K.~L.}\ \bibnamefont {He}},\ }\href {\doibase 10.1088/1674-1137/36/8/010} {\bibfield  {journal} {\bibinfo  {journal} {Chin. Phys. C}\ }\textbf {\bibinfo {volume} {36}},\ \bibinfo {pages} {742} (\bibinfo {year} {2012})}\BibitemShut {NoStop}%
\bibitem [{\citenamefont {Petri}(2010)}]{Petri:2010ea}%
  \BibitemOpen
  \bibfield  {author} {\bibinfo {author} {\bibfnamefont {T.}~\bibnamefont {Petri}},\ }\emph {\bibinfo {title} {{Anomalous decays of pseudoscalar mesons}}},\ \href@noop {} {\bibinfo {type} {Other thesis}} (\bibinfo {year} {2010}),\ \Eprint {http://arxiv.org/abs/1010.2378} {arXiv:1010.2378} \BibitemShut {NoStop}%
\bibitem [{\citenamefont {James}\ and\ \citenamefont {Roos}(1975)}]{James:1975dr}%
  \BibitemOpen
  \bibfield  {author} {\bibinfo {author} {\bibfnamefont {F.}~\bibnamefont {James}}\ and\ \bibinfo {author} {\bibfnamefont {M.}~\bibnamefont {Roos}},\ }\href {\doibase 10.1016/0010-4655(75)90039-9} {\bibfield  {journal} {\bibinfo  {journal} {Comput. Phys. Commun.}\ }\textbf {\bibinfo {volume} {10}},\ \bibinfo {pages} {343} (\bibinfo {year} {1975})}\BibitemShut {NoStop}%
\bibitem [{\citenamefont {Langenbruch}(2022)}]{Langenbruch:2019nwe}%
  \BibitemOpen
  \bibfield  {author} {\bibinfo {author} {\bibfnamefont {C.}~\bibnamefont {Langenbruch}},\ }\href {\doibase 10.1140/epjc/s10052-022-10254-8} {\bibfield  {journal} {\bibinfo  {journal} {Eur. Phys. J. C}\ }\textbf {\bibinfo {volume} {82}},\ \bibinfo {pages} {393} (\bibinfo {year} {2022})}\BibitemShut {NoStop}%
\bibitem [{\citenamefont {Ablikim}\ \emph {et~al.}(2013)\citenamefont {Ablikim} \emph {et~al.}}]{BESIII:2012mpj}%
  \BibitemOpen
  \bibfield  {author} {\bibinfo {author} {\bibfnamefont {M.}~\bibnamefont {Ablikim}} \emph {et~al.} (\bibinfo {collaboration} {BESIII Collaboration}),\ }\href {\doibase 10.1103/PhysRevD.87.012002} {\bibfield  {journal} {\bibinfo  {journal} {Phys. Rev. D}\ }\textbf {\bibinfo {volume} {87}},\ \bibinfo {pages} {012002} (\bibinfo {year} {2013})}\BibitemShut {NoStop}%
\bibitem [{\citenamefont {Bickert}\ and\ \citenamefont {Scherer}(2020)}]{Bickert:2020kbn}%
  \BibitemOpen
  \bibfield  {author} {\bibinfo {author} {\bibfnamefont {P.}~\bibnamefont {Bickert}}\ and\ \bibinfo {author} {\bibfnamefont {S.}~\bibnamefont {Scherer}},\ }\href {\doibase 10.1103/PhysRevD.102.074019} {\bibfield  {journal} {\bibinfo  {journal} {Phys. Rev. D}\ }\textbf {\bibinfo {volume} {102}},\ \bibinfo {pages} {074019} (\bibinfo {year} {2020})}\BibitemShut {NoStop}%
\bibitem [{\citenamefont {Escribano}\ \emph {et~al.}(2015)\citenamefont {Escribano}, \citenamefont {Masjuan},\ and\ \citenamefont {Sanchez-Puertas}}]{Escribano:2015nra}%
  \BibitemOpen
  \bibfield  {author} {\bibinfo {author} {\bibfnamefont {R.}~\bibnamefont {Escribano}}, \bibinfo {author} {\bibfnamefont {P.}~\bibnamefont {Masjuan}}, \ and\ \bibinfo {author} {\bibfnamefont {P.}~\bibnamefont {Sanchez-Puertas}},\ }\href {\doibase 10.1140/epjc/s10052-015-3642-z} {\bibfield  {journal} {\bibinfo  {journal} {Eur. Phys. J. C}\ }\textbf {\bibinfo {volume} {75}},\ \bibinfo {pages} {414} (\bibinfo {year} {2015})}\BibitemShut {NoStop}%
\end{thebibliography}%

\end{document}